\documentclass[12pt, draftclsnofoot, onecolumn]{IEEEtran}

\usepackage{lineno}
\usepackage{tabularx}
\usepackage{makecell}
\usepackage{amsmath}
\usepackage{graphicx}
\usepackage{subfigure}
\usepackage{multicol} 
\usepackage{epstopdf} 
\usepackage{epsfig} 
\usepackage{multirow}
\usepackage{bm}
\usepackage{array}
\usepackage{color}
\usepackage{float}
\usepackage{amsfonts}
\usepackage{caption}
\usepackage{chem formula}

\AtBeginDocument{\nolinenumbers}
\begin{document}
%\linenumbers
%\title{Unraveling the fundamental role of dendritic sublinearity in neuronal information processing }
\title{Implementing feature binding through dendritic networks of a single neuron}

\author{\IEEEauthorblockN{Yuanhong Tang\dag, Shanshan Jia\dag, Tiejun Huang\dag, Zhaofei~Yu\dag, Jian~K.~Liu\ddag \\}
	\small{\IEEEauthorblockA{
	\dag  School of Computer Science, Institute for Artificial Intelligence, Peking University, Beijing, China\\
	\ddag School of Computer Science, Centre for Human Brain Health, University of Birmingham, Birmingham, UK \\
		%Correspondence:: yuzf12@pku.edu.cn; j.liu.22@bham.ac.uk 
        }
        }
}

\maketitle

\begin{abstract}

A single neuron receives an extensive array of synaptic inputs through its dendrites, raising the fundamental question of how these inputs undergo integration and summation, culminating in the initiation of spikes in the soma. Experimental and computational investigations have revealed various modes of integration operations that include linear, superlinear, and sublinear summation. Interestingly, different types of neurons exhibit diverse patterns of dendritic integration depending on the spatial distribution of dendrites. The functional implications of these specific integration modalities remain largely unexplored. In this study, we employ the Purkinje cell (PC) as a model system to investigate these complex questions. Our findings reveal that PCs generally exhibit sublinear summation across their expansive dendrites. Both spatial and temporal input dynamically modulates the degree of sublinearity. Strong sublinearity necessitates the synaptic distribution in PCs to be globally scattered sensitive, whereas weak sublinearity facilitates the generation of complex firing patterns in PCs. Using dendritic branches characterized by strong sublinearity as computational units, we demonstrate that a neuron can successfully address the feature binding problem. Taken together, these results offer a systematic perspective on the functional role of dendritic sublinearity, inspiring a broader understanding of dendritic integration in various neuronal types.

\end{abstract}

\section{Introduction}

Neurons, the foundational computational entities within the complex brain system, precisely process signal input through their dendritic spine. The conventional understanding of dendritic function posits that these structures receive synaptic input and subsequently convey it to the soma, thus generating action potentials or spikes. This proposition suggests that a single neuron has computational capabilities to execute information processing through an extensive array of dendritic branches. However, the underlying integrating principles governing the diverse array of dendrites and neuron types remain elusive. In recent decades, experimental studies have facilitated laborious explorations of the integration of synaptic inputs along the dendritic domains of individual neurons, shedding light on the physiological basis of dendrite function~\cite{davie2006dendritic, spruston2008pyramidal, spruston2016principles, lafourcade2022differential}. Dendrites have been shown to function as autonomous information processing units, capable of performing local computations, thus conferring substantial computational capabilities on single neurons~\cite{branco2010single, london2005dendritic, koch2000role, poirazi2020illuminating, stuart2015dendritic, makarov2023dendrites}. Furthermore, these insights have contributed significantly to our understanding of how neurons engage in communication with each other and the resultant impact on behavior~\cite{cash1999linear, polsky2004computational, larkum1999new, branco2010dendritic}.

The principal function of dendrites lies in the sophisticated integration and processing of synaptic inputs~\cite{london2005dendritic, major2013active}. Investigations focusing on pyramidal cells reveal that dendritic integration is characterized by superlinear dynamics, attributed to the activation of voltage-dependent channels and NMDARs~\cite{polsky2004computational, branco2011synaptic, losonczy2006integrative, gasparini2006state}. In particular, this superlinear integration has been observed primarily in cortical neurons~\cite{cash1999linear, hu2010dendritic, carter2007timing}, while sublinear summation has been documented in cerebellar interneurons ~\cite{abrahamsson2012thin, tran2016differential}. The manifestation of dendritic nonlinearity allows neurons to function as feature detectors~\cite{branco2010dendritic, larkum1999new}, exerting influence on various aspects of brain function, including whisker sensation~\cite{xu2012nonlinear}, orientation selectivity~\cite{wilson2016orientation}, sensory perception~\cite{takahashi2016active, lavzin2012nonlinear}, sensory-motor integration~\cite{xu2012nonlinear}, and memory encoding~\cite{kaifosh2016mnemonic, tzilivaki2019challenging}. Furthermore, nonlinear dendritic integration is postulated to augment the computational capacity of neurons~\cite{katz2009synapse, koch1983nonlinear}, enabling them to compute linearly non-separable functions~\cite{caze2013passive, caze2023demonstration}. The incorporation of nonlinear dendritic mechanisms into artificial neural networks is proposed to bring neurons in comparative alignment with their biological counterparts~\cite{poirazi2003pyramidal, beniaguev2021single, tzilivaki2019challenging, branco2010single}. The benefit of installing dendritic nonlinearity into artificial neural networks not only reduces power consumption and enhances accuracy~\cite{li2020power, chavlis2021drawing} but also mitigates communication costs within neural networks~\cite{wu2023mitigating}.

Most investigations of dendritic nonlinearity have mainly focused on cortical neurons. However, there is a substantial gap in our understanding of the dendritic nonlinearity inherent in the most complex neuronal entities, namely cerebellar Purkinje cells (PCs)~\cite{galakhova2022evolution}. The complex dendritic structures of PCs contribute to the amplification of the computational repertoire within single neurons~\cite{anCodingCapacityPurkinje2019}. PCs, characterized by receiving thousands of synaptic inputs in a location-dependent manner, exhibit a distinctive synaptic organization, with the distal region of the dendritic tree receiving excitatory input from parallel fibers originating from granule cells~\cite{napper1988number, tangRegulatingSynchronousOscillations2021}. On the contrary, the proximal portion of the dendritic tree is innervated by the input of climbing fibers~\cite{watanabe2011climbing}. The interplay between synaptic input and dendritic nonlinear properties produces highly precise input-output functions~\cite{beniaguev2021single}. Therefore, it is imperative to understand whether passive cable properties contribute to location-dependent dendritic integration and whether dendritic synaptic integration plays a role in PC computation.

In this study, we employ detailed PC models that capture the intricacies of dendritic morphology to investigate dendritic nonlinear synaptic integration and its implications in information processing. Our examination of subthreshold synaptic summation in PCs reveals a significant sublinear synaptic integration profile. Notably, the strong sublinearity of spiny dendrites requires PCs to have globally scattered input to fire somatic spikes.  
Furthermore, we observe that weak sublinear responses in smooth dendrites promote the generation of bursts, potentially leading to the generation of complex spikes.  From the perspective of dendritic nonlinearity, this explains why the synaptic distribution in PCs is location-dependent. Importantly, this sublinearity enables a single neuron to compute feature binding problems. Our results suggest that while nonlinear dendrites may not supplant neurons as the fundamental computational units, they undoubtedly expand the computational capacities of neuronal systems.

\section{Methods}

\subsection{Neuronal model}
In this study, we used a detailed neuron model incorporating neuromorphic structures to explore the dendritic nonlinear synaptic integration capabilities of Purkinje cells and how dendritic nonlinearity affects the neuronal response and computational capabilities. This model extends the traditional Hodgkin-Huxley framework~\cite{hodgkinQuantitativeDescriptionMembrane1952} by including detailed representations of the neuron's morphology, such as dendrites and soma compartments, to better capture the spatial and temporal dynamics of electrical signaling. The model describes the membrane potential $ V$ through a set of nonlinear differential equations. These equations account for the ionic currents across the neuronal membrane and leakage currents. In the neuromorphic model, these equations are applied to multiple compartments representing different parts of the neuron. The membrane potential dynamics for each compartment \textit{$ i$} are given by:
\begin{equation}
\begin{split}
C_{m,i} \frac{dV_i}{dt} = -I_{\text{ion},i} - I_{L,i} - I_{\text{syn},i} + \sum_j g_{ij}(V_j - V_i),
\end{split}
\end{equation}
 where $C_{m,i}$ is the membrane capacitance of compartment $i$, $ I_{\text{ion},i}$  is the ion current in compartment $ i$, $ I_{L,i}$  is the leakage current in compartment $ i$, $ I_{\text{syn},i} $ is the synaptic current applied to compartment $ i$, $ V_{j}$ is the membrane potential of compartment $j$, $ g_{ij}$  is the conductance between compartments $ i$ and $ j$:

 \begin{equation}
\begin{split}
g_{ij} = \frac{\pi d_i^2 d_j^2}{2 R_i \left( L_i d_j^2 + L_j d_i^2 \right)},
\end{split}
\end{equation}
where $d_{i,j}$ and $L_{i,j}$ are the diameters and lengths of compartments $i$ and $j$, respectively. $R_i$ is the axial resistance.

Ion channels are fundamental components of the neuronal membrane and are responsible for generating and propagating electrical signals in neurons. The ion current $ I_{\text{ion}} $ is given by:
\begin{equation}
\begin{split}
I_{\text{ion}} = g_{\text{ion}} m^n h^i (V - E_{\text{ion}}),
\end{split}
\end{equation}
where $ g_{\text{ion}} $ is the maximum ion conductance, $ m $ is the activation gating variable,  $ h$ is the inactivation gating variable, $ n$ and $ i$ represent the number of activation  and inactivation gates respectively, $ V$ is the membrane potential, and $ E_{\text{ion}}$ is the ion reversal potential. The gating variables $ m $ and $ h $ follow first-order kinetics with voltage-dependent rate constants:
\begin{equation}
\begin{split}
\frac{dm}{dt} = \alpha_m (1 - m) - \beta_m m,\\
\frac{dh}{dt} = \alpha_h (1 - h) - \beta_h h, 
\end{split}
\end{equation}
where $ \alpha_{m,h}$ and $ \beta_{m,h}$  are rate constants for the ion channel. The leakage ion channel represents the passive flow of ions through the membrane, contributing to the resting membrane potential. The leakage current $ I_{L}$ is given by:

\begin{equation}
\begin{split}
I_{L} &= g_{L}(V - E_{L}),\\
g_{L} &=1/R_m,
\end{split}
\end{equation} 
where $ g_{L}$  is the leakage conductance, $ V$ is the membrane potential, $ E_L$ is the leakage reversal potential, $R_m$ is the membrane resistance. 
 
We used a 3D reconstruction of a mouse PC (NMO-00865) available on the public archive (www.neuromorpho.org). The model contains a somatic compartment and 739 dendrites. The passive properties and voltage-dependent ionic channels used in this study are consistent with those reported in~\cite{tang2023diverse}, which investigates the role of NMDA receptors across different dendritic branches and regions of Purkinje cells. PC model parameters were primarily derived from~\cite{de1994active}, with additional ionic channel parameters were adopted from~\cite{miyasho2001low,marasco2013using}.  $ R_m$ and  $ R_i$ were set to 5000 $ \Omega *c{m^2}$ and 250 $\Omega *cm$, respectively. Membrane capacitance ($ C_m$)  was set to 0.8 $ \mu F/c{m^2}$ in the soma and main dendrites, and 0.5 $ \mu F/c{m^2}$ in spiny dendrites. 13 diﬀerent types of voltage-gated ion channels were installed in the PC model. Three channels (fast $ Na^+$ channel (NaF), persistent $ Na^+$ channel (NaP), anomalous rectifier channel (Kh)) were solely added to the soma, and two channels (high-threshold $ Ca^{2+}$-activated ${ K^+}$channel (KC3), low-threshold $ Ca^{2+}$-activated ${ K^+}$ channel (K23)) were solely added to the dendrites. Eight (P-type $ C{a^{2 + }}$ channel (CaP), T-type $ Ca^{2+}$ channel (CaT), class-E $ Ca^{2+}$ channel (CaE), persistent $ {K^+}$ channel (KM), A-type $ {K^+}$ channel (KA), D-type $ {K^ + }$ channel (KD), delayed rectifier (Khh), decay of sub-membrane $ Ca^{2+}$) were inserted into the soma and dendrites. The decay of sub-membrane $ Ca^{2+}$ can be described as:
\begin{equation}
\begin{split}
&\frac{d[\text{Ca}^{2+}]_i}{dt} = \text{drive\_channel} + \text{drive\_pump} + \frac{[\text{Ca}^{2+}]_\infty - [\text{Ca}^{2+}]_i}{\tau_r},\\
&\text{drive\_channel} = -\frac{10000 \cdot I_{Ca}}{2 \cdot F \cdot \text{D}},\\
&\text{drive\_{pump}} = - \frac{k_t \cdot [\text{Ca}^{2+}]_i}{[\text{Ca}^{2+}]_i + k_d},
\end{split}
\end{equation} 
where the steady-state calcium concentration $[\text{Ca}^{2+}]_\infty$ is $4e^{-5}$ mM, the calcium removal time constant $\tau_r$ is 2ms, the Faraday constant (F) is 96,489, the depth of the shell (D) is $0.1 \mu m$, the maximum rate of the pump $K_t$ is $4e^{-5}$ mM/ms, and the dissociation constant of the pump $k_d$ is  $4e^{-5}$ mM. The parameters for the remaining ion channels are listed in Table ~\ref{table1}.
\begin{table}[h!]
\centering
\caption{Ion channel parameters.}
\begin{tabularx}{\textwidth}{|c|X|X|X|X|}
\hline
\textbf{Ion Channel} & \textbf{\makecell{Maximum \\ Conductance \\ (g\_ion, nS)}} & \textbf{\makecell{Activation \\ Gate (n)}} & \textbf{\makecell{Inactivation \\ Gate (i)}} & \textbf{\makecell{Reversal \\ Potential \\ (E\_ion, mV)}} \\ \hline
NaF & 10 & 3 & 1 & 45 \\ \hline
NaP & 0.001 & 3 & \textbackslash & 45 \\ \hline
Kh & 0.0005 & 1 & \textbackslash & -30 \\ \hline
KC3 & 0.06 & 1 & 2 & -85 \\ \hline
K23 & 0.00039 & 1 & 2 & -85 \\ \hline
CaP & 0.004 & 1 & \textbackslash & 135 \\ \hline
CaI & 0.0015 & 1 & 1 & 135 \\ \hline
CaE & 0.008 & 1 & 1 & 135 \\ \hline
KM & 0.00001 & 1 & \textbackslash & -85 \\ \hline
KA & 0.08 & 4 & 1 & -85 \\ \hline
KD & 0.09 & 1 & 1 & -85 \\ \hline
Khh & 0.0006 & 4 & \textbackslash & -85 \\ \hline
\end{tabularx}
\label{table1}
\end{table}

\subsection{Synapse model}
Synapses are critical structures that mediate communication between neurons through the release of neurotransmitters. Synapse models are based on the biophysical properties of synaptic currents and their interaction with the postsynaptic membrane.  To explore the dendritic nonlinear synaptic integration ability of PCs, two excitatory synapse models, AMPA and NMDA, were included. Excitatory synapses typically involve the release of neurotransmitters such as glutamate, which bind to receptors on the postsynaptic neuron, leading to depolarization. The excitatory postsynaptic current (EPSC) can be  represented as:
\begin{equation}
\begin{split}
I_{syn}=g_{syn}*(V-E_{syn}),\\
g_{syn}=g_{max}*w*Y*S(t),  
\end{split}
\end{equation} 
where $ {I_{syn}}$ is the receptor current, $ {g_{syn}}$ is the receptor conductance,  $ V$ is the membrane potential, $ {{E_{syn}}}$ is the receptor reversal potential, $ {g_{\max}}$ is the maximum synaptic conductance, and $ w$ is the connection weight. 

NMDA-mediated currents typically require AMPA receptor-mediated depolarization to relieve the extracellular $ Mg^{2+}$ blockade on NMDA-associated channels. These channels only open when the magnesium ions are displaced, allowing NMDA receptors to be activated concurrently. The scaling factor $Y$ is 1 for AMPA, while it is a nonlinear voltage ($V$) dependent function for NMDA: $Y= 1 / (1 + 0.24*exp(-154*V/0.027)).$
%\begin{equation}
%\begin{split}
%\left\{
%\begin{array}{l}
%Y= 1 / (1 + 0.24*exp(-154*V/0.027))~for~NMDA, \\
%Y=1~for~AMPA.
%\end{array}
%\right.
%\end{split}
%\end{equation}

The synaptic gating variable $S(t)$ follows double exponential kinetics:
\begin{equation}
\begin{split}
S(t) =  e^{-(t - t_{spike})/\tau_{decay}} - e^{-(t - t_{spike})/\tau_{rise}},
\end{split}
\end{equation}
where  $ \tau_{rise} $ is the rise time constant,  $ \tau_{decay}$ is the decay time constant, and $ t_{spike}$ represents the time of the presynaptic spike. $ \tau_{rise}$  was set to 0.5 ms for AMPA and 8 ms for NMDA. $\tau_{decay}$ was set to 5 ms for AMPA and 30 ms for NMDA~\cite{de1994active, ozawa1998glutamate}. When the excitatory synapse includes both an AMPA and an NMDA component, they are co-located and always co-activated.  When calculating dendritic nonlinearity, synapses are distributed starting at the 0.1 position of the dendrite, then added distally at intervals of  $ d=0.2$ from the starting point. 

Each synapse receives an independent spike train as stimulus input. A single stimulation consists of a sequence of spikes, including their spike times and inter-spike intervals. Consequently, a successive spike train can be generated by adding regular or irregular time intervals to the previous spike. 
We utilized a Poisson-distributed pulse sequence model to investigate the response of neurons to stochastic external stimuli. 
The spike train was generated according to a Poisson process, characterized by an average firing rate $ \lambda$ (pulses per second). The intervals between pulses $ \Delta t$ were exponentially distributed with parameter $ \lambda$, ensuring that the pulse arrivals are memoryless and randomly timed. The probability density function of the inter-pulse intervals is given by:
\begin{equation}
\begin{split}
P(\Delta t) = \lambda e^{-\lambda \Delta t},
\end{split}
\end{equation}
where  $ \Delta t$ is the time between consecutive pulses. Firstly, we generate a random number  $ u $ uniformly distributed in the interval \((0, 1)\). Then compute the inter-pulse interval  $ \Delta t$ using the inverse transform sampling method:
\begin{equation} 
\begin{split}
\Delta t = -\frac{1}{\lambda} \ln(1 - u),
\end{split}
\end{equation}
and the next pulse arrival time is $ t_{i+1} = t_i + \Delta t$, and this process was repeated to generate a sequence of pulse arrival times over the desired simulation period.

In this study, we examined the effects of both synchronous and asynchronous inputs on neuronal activity.  Synchronous inputs involve multiple synapses receiving input pulses simultaneously which can lead to a strong and coordinated response in the postsynaptic neuron. Asynchronous inputs involve multiple synapses receiving input pulses at different times, leading to a more distributed and less coordinated response in the postsynaptic neuron.

\subsection{Data analysis}
We simulated the PC morphology model in NEURON 8.0. The time step of all simulation experiments was set at 0.025 ms.  Data collected after simulations were saved for further analysis.
In the default model, we turned off ionic currents in simulations. By removing active ion channel distributions, we aimed to simplify the model and focus on how the spatial arrangement of synapses and the inherent passive properties of dendrites shape integration, free from the influence of active currents. However, the dendritic nonlinearity is even more pronounced compared to the passive model, when active channels are turned on. To focus on the effect of nonlinearity on PC dynamics, we used the passive model as the default model.     
%In order to avoid the influence of ionic current on the nonlinear synaptic integration ability of dendrites, there is no ion channel distribution in all compartments of PC when calculating dendritic nonlinearity. 

To study dendritic nonlinearity, we record the observed EPSPs ($ EPSP_O$), and the EPSPs produced by individual synapses are added together to obtain the expected EPSP ($ EPSP_E$). The sublinearity index was measured as the average sublinear summation $ (1- (EPSP_O/EPSP_E)) * 100$ with the last two points as in the dashed box in Figure 1. To investigate the temporal dependence of dendritic integration, we systematically varied the inter-spike intervals (ISI) between synaptic inputs. Then, the amplitude of all EPSPs induced by the stimuli was averaged to obtain the EPSP value elicited by the stimulus. From this, the observed values and expected values were obtained, and the sublinearity index was calculated. 

In addition, the expected and observed EPSP curves are fitted by the Hill function as previously~\cite{rothman2009synaptic}: 
\begin{equation}
 F (EPSP_E) =\frac{EPSP_O{max}}{1+(EPSP_E{50}/EPSP_E)^n}+EPSP_O{0},
\end{equation}
where $ n$ is the exponent factor, $ EPSP_O{0}$ is the observed EPSP offset, and  $ EPSP_O{max}$ is the maximum observed EPSP. $ EPSP_E{50}$ is the value of expected EPSP at which observed EPSP reaches half maximum. To characterize EPSP in detail, we introduced three measures, the peak, rise time, and width of EPSP. Peak is the maximum peak value of EPSP, and width as the distance between the points where the half-amplitude reaches before and after the peak.

To further study the effect of dendritic nonlinearity on PC firing. The dendrites with less than $20\%$ sublinearity are regarded as main dendrites, and those with more than $20\%$ are regarded as spiny dendrites, and then the spiny dendrites are divided into 24 branches, and the parent node of each branch is the main dendrite.

\subsection{Implementation of Feature binding problem} 

Feature binding problem (FBP) refers to the brain's ability to integrate multiple attributes (e.g., color, shape, motion) into a unified percept.  In the context of Purkinje cell computation, each dendritic branch is treated as an independent computational unit. Synaptic activation on a branch is represented as input as 1, while the absence of activation is represented as 0. In the simplest case, 2 out of 24 dendritic branches are chosen as inputs ($X_1, X_2$). These inputs represent binary features, and the PC was tasked with performing basic Boolean operations, such as AND, OR, or XOR. This level tests the PC's ability to compute simple logical functions based on spatially distributed inputs. Specific dendritic branches were selected to represent distinct features. In the spatial FBP task, any 4 out of 24 branches were designated as $X_1, X_2, X_3, X_4$, where the neuron is required to bind these features into a coherent response. The neuron's response depends not only on which branches are activated but also on the specific combination of inputs.

To incorporate temporal dynamics, we extended the FBP to a spatiotemporal setting. Here, the inputs ($X_1, X_2, X_3, X_4$) are presented in three sequential temporal steps ($t_1, t_2, t_3$). Two features are activated in each step, and the neuron's response depends not only on the spatial pattern of inputs but also on their timing. The neuron fires only when $X_3$ and $X_4$ are activated during the second $t_2$ and third $t_3$ steps. Any change in the temporal order or timing of activation prevents PC firing. In this way, PC is dedicated to select the spatiotemporal patterns of assigned features. 

% Feature binding refers to the brain's ability to integrate various attributes (e.g., color, shape, motion) into a coherent percept. This process involves complex, nonlinear interactions within neuronal dendrites. When there are multiple synaptic inputs $ I_1, I_2, \ldots, I_n$. The nonlinear response  $ R$  can be described by a function $ f$:
% \begin{equation}
% \begin{split}
% R = f(I_1 + I_2 + \ldots + I_n),
% \end{split}
% \end{equation}
% When studying PC computing performance, each branch is considered a computing unit. When the synapses distributed on the branch are activated, the input $ I_n$ is considered to be 1, and when it is not activated, the input $ I_n$ is considered to be 0. When studying Boolean functions, choose any 2 from 24 branches as $ X_1$ and $ X_2$. when studying the feature binding problem, 4 of the 24 branches are selected as $ X_1$, $ X_2$, $ X_3$, and $ X_4$.

\section{Result}
  
\subsection{PC dendrites show sublinear integration}

 %Cable theory and experiments predict a decrement in the amplitude of somatic voltage with synaptic activation at increasing distances from the soma.  
  \begin{figure*}[tbp]
	\centering	
    \includegraphics[width=1\textwidth]{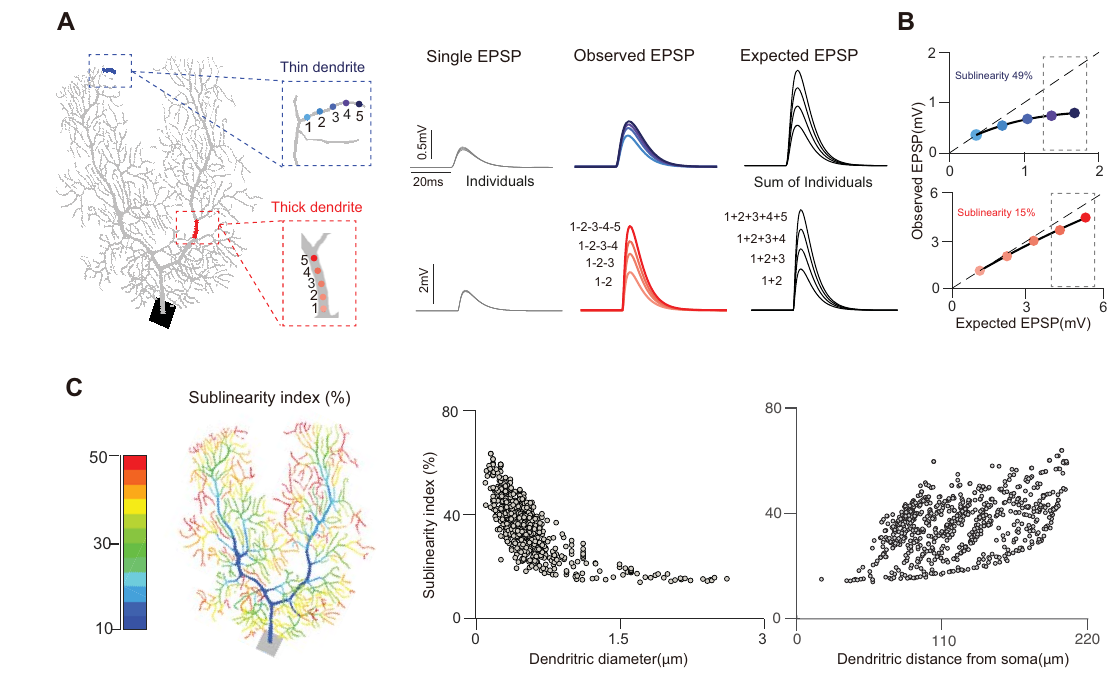} 
	\caption{ PC individual dendrites show sublinear integration to the somatic response.
    (A) (Left) 5 synapses are distributed on thin (blue) and thick (red) dendrites. (Right) Single EPSPs evoked by single-pulse synaptic stimulation; Observed EPSPs evoked with increasing synapses; Expected EPSPs as the algebraic sum of single EPSPs. 
    (B) Sublinear relationship of the expected and observed EPSPs from (A). The dotted line represents a linear function. The sublinearity index, 49\% (thin) and 15\% (thick) was measured as the average sublinear summation in the dashed box. The solid line indicates fits by a Hill function.
    (C) The distributed sublinearity index for each dendrite. Sublinearity as a function of dendritic diameter and distance from soma.
	}
	\centering
	\label{fig1}
\end{figure*}

 \begin{figure*}[tbp]
	\centering	
    \includegraphics[width=1\textwidth]{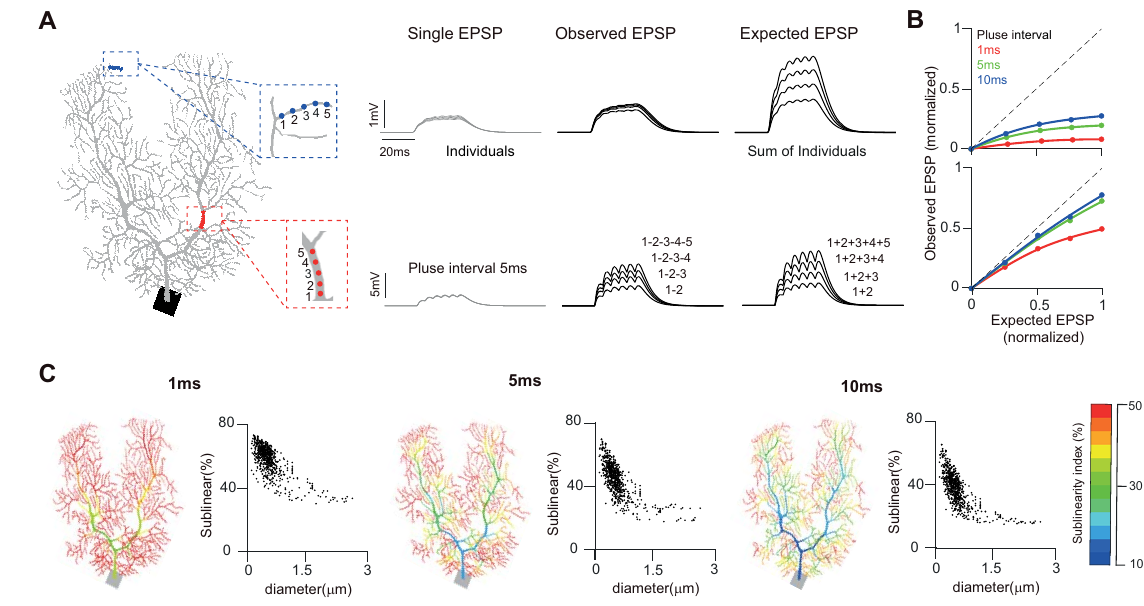}
	\caption{ Temporal dependence of sublinear dendritic integration.
    (A) Synaptic inputs are distributed across thin (blue) and thick (red) dendritic (left). Single EPSPs were evoked using six-pulse synaptic stimulation, and the observed EPSPs were measured with an increasing number of active synapses. The expected EPSPs were calculated as the algebraic sum of the individual EPSPs (right). A pulse interval of 5 ms was used.
    (B) The relationship between observed and expected EPSPs at pulse intervals of 1 ms, 5 ms, and 10 ms, normalized to the maximum response. The dotted line indicates the expected linear summation, while the solid lines represent fits to the Hill function.
    (C) The sublinearity index is mapped across the dendritic for pulse intervals of 1 ms, 5 ms, and 10 ms, with the color scale indicating the degree of sublinearity (left). Sublinearity is plotted as a function of dendritic diameter (right).    
	}
	\centering
	\label{fig2}
\end{figure*}

To examine whether the large dendritic depolarization could decrease the synaptic current driving force and introduce a nonlinearity that would curtail linear summation of somatic EPSPs (excitatory postsynaptic potentials) within the same PC dendrite, we studied subthreshold input-output relationships by comparing the algebraic sum of individual EPSP from five activated synapses along different yet nearby spatial locations along one dendrite, with the compound EPSP in response to simultaneous activation of multiple 2 to 5 locations (Figure~\ref{fig1}A). The compound EPSP observed ($ {EPSP_O}$) was systematically smaller than the algebraic sum of its corresponding individual EPSP expected ($ {EPSP_E}$), especially on thin dendrites. The input-output relationship of dendritic summation using these five synapses shows sublinear trends that can be fitted by the Hill function. The sublinearity index (SI) can be calculated (the average of $ {(1-( EPSP_O/ EPSP_E))*100}$ using the last two points; dashed box in Figure~\ref{fig1}B) for each dendrite. The SI of the thin dendrite (diameter $0.29 \mu m$, 49\%) is stronger than the thick dendrite (diameter $1.75 \mu m$, 15\%) as in Figure~\ref{fig1}B. We then asked how dendritic properties regulate the SI (Figure~\ref{fig1}C). The SI of 739 dendrites was distributed over the entire dendritic field, which shows apical dendrites exhibit stronger SIs, but trunk dendrites are weaker.  SI decreases with larger diameter dendrites while increasing with the dendritic distance from the soma. Moreover, SI increased significantly with the addition of active ion channels installed on the PC model, indicating that active ion channel dynamics enhance dendritic sublinearity (Figure~\ref{suppfig1}). Such observations are inherent in the observed and expected EPSPs (Figure~\ref{suppfig2}). The peak, rise time, and width of EPSP are also location-dependent (Figure~\ref{suppfig3}). We further observed consistent nonlinearity patterns with several additional PCs with different  morphologies from various species (Figure~\ref{suppfig4}), including a rat PC (NMO-00891), a mouse PC (NMO-00864), and a guinea pig PC (NMO-00610) available on public archives (www.neuromorpho.org). %Across all these morphologies, we observed consistent patterns that dendritic sublinearity is stronger in thin dendrites and weaker in thicker dendrites.

To further explore the effect of temporal summation on a sequence of synaptic inputs on dendritic sublinearity, we used stimulation protocols with six temporal pluses at different intervals (1, 5, and 10 ms) with the same five spatial synaptic locations (Figure~\ref{fig2}A). Indeed the SI can be significantly changed by temporal intervals (Figure~\ref{fig2}B) for individual dendrites. The smaller interval can increase SI, while the larger interval reduces SI. Such a change is systematically preserved over all the dendrites  (Figure~\ref{fig2}C). For simplicity, the 5 ms interval was used throughout this study. These results suggest that PC dendritic summation shows strong sublinearity.

\subsection{Strong sublinearity prefers a global scatter input strategy}

  \begin{figure*}[tbp]
	\centering	
    \includegraphics[width=1\textwidth]{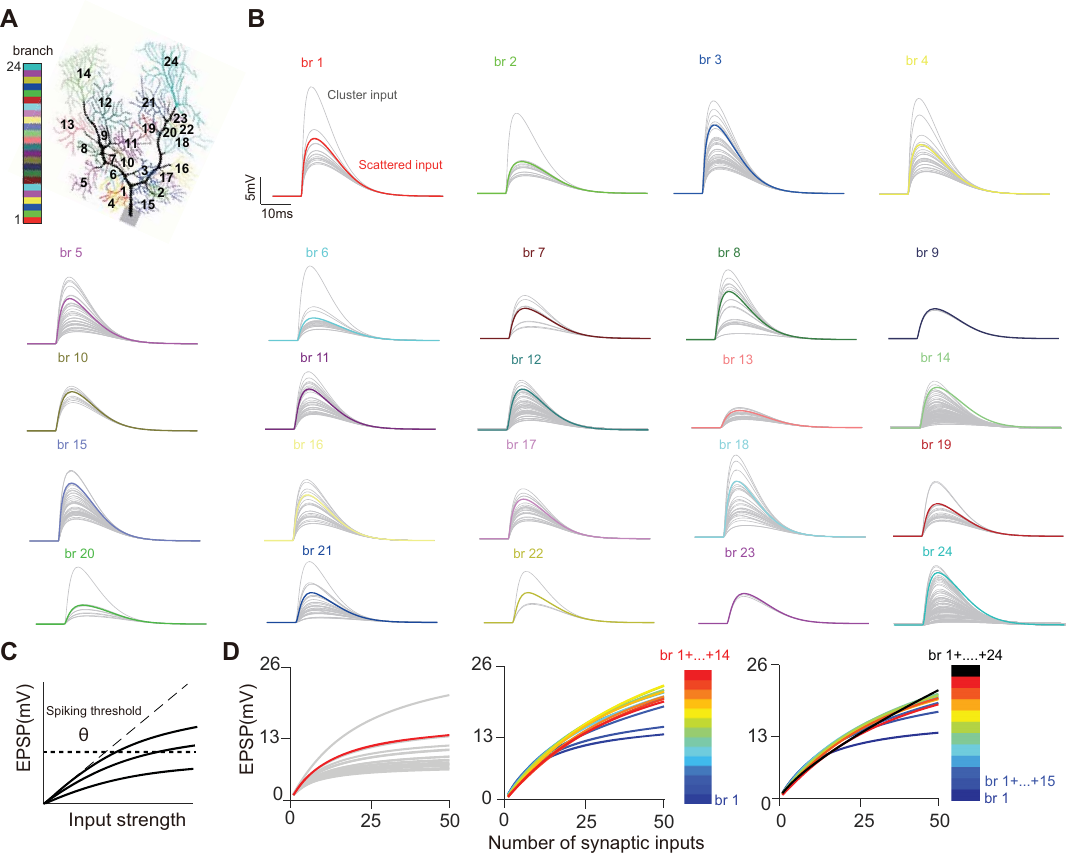}
	\caption{ Scattered synaptic inputs on strong sublinear dendrites trigger larger somatic responses. 
  (A) 24 branches (color-coded) with strong sublinearity are segmented in the entire morphology. Each branch groups a set of single dendrites with a sublinearity index >20\%. The black trunk represents the main dendrites (dendritic sublinearity <20\%). 
  (B) Somatic EPSP in response to 50 synaptic inputs in a branch. Individual gray traces represent the response to a cluster input (50 synapses are distributed closely) on a single dendrite. The color trace shows synapses scattered throughout the branch. 
  (C) Illustration of the somatic spiking threshold $\theta$ with varying degrees (solid lines) of dendritic sublinearity. 
  %Sub-threshold input/output relationships used to quantify dendritic operations. The dashed line represents a linear relationship. $\theta$ indicate example somatic spike thresholds. Three solid lines indicate varying degrees of sublinearity. 
  Strong sublinearity makes soma voltage more difficult to reach the spiking threshold. 
  (D) Input-output (IO) relationship curves change over the sum of dendritic branches. (Left) IO curves in branch 1 (br1 in (B)). Gray traces represent cluster input on single dendrites. The color trace shows synapses scattered throughout the branch.  (Right) IO curves were obtained by scattered synaptic inputs throughout the cumulative branches on top of br1. 
	}
	\centering
	\label{fig3}
\end{figure*}

   \begin{figure*}[tbp]
	\centering	
    \includegraphics[width=1\textwidth]{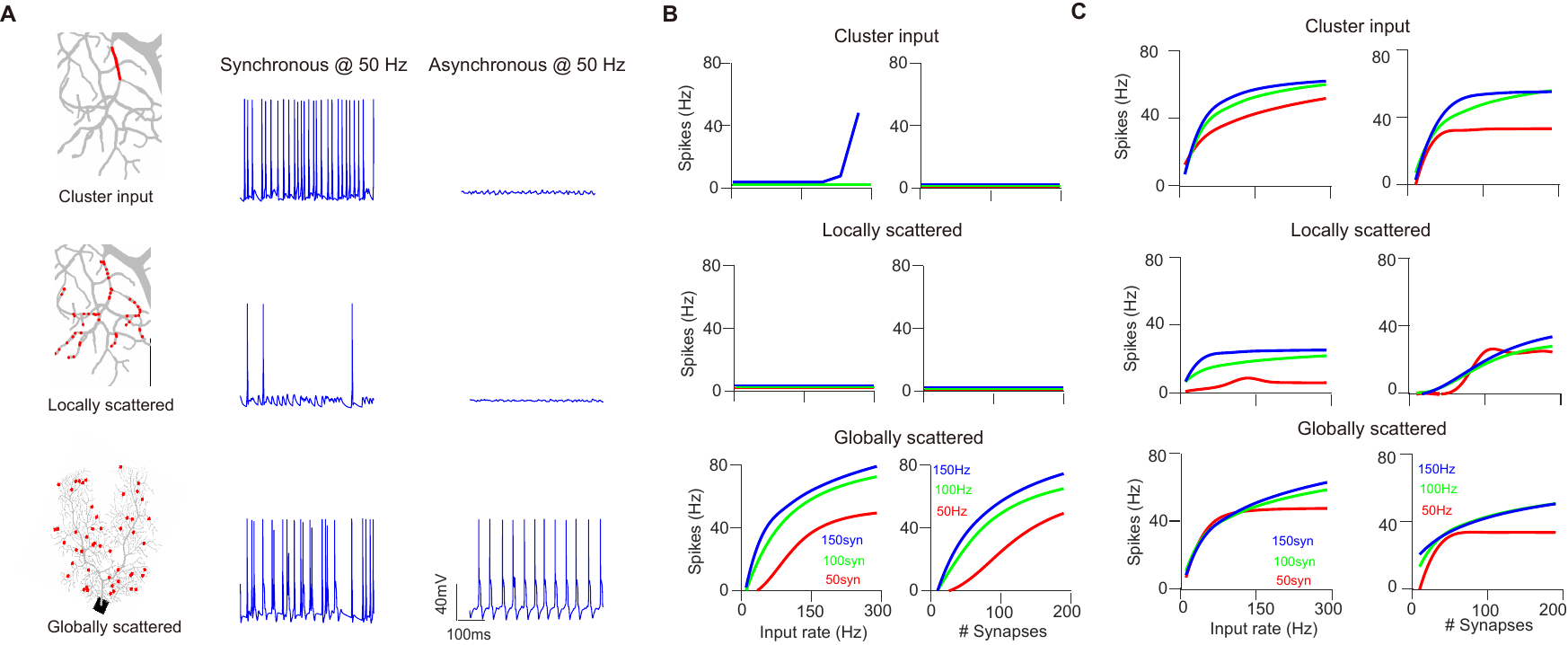}
	\caption{ Somatic spiking response boosted by globally scattered synaptic inputs across strong sublinear dendrites.
   (A) Spiking activity generated by synaptic inputs under three spatial and two temporal patterns. (Left) cluster input: synapses distributed in a single dendrite. (Middle) locally scattered: synapses distributed in individual branches. (Right) globally scattered: synapses distributed across all branches. Somatic spiking activity in response to either synchronous or asynchronous 50 synaptic inputs under each spatial stimulation pattern at 50 Hz.
   (B) Spiking activity vs. asynchronous input under three spatial stimulation patterns with (left) 50, 100, and 150 synapses; (right) 50, 100, and 150 Hz. 
(C). Similar to B, but with synchronous input. 
	} 
	\centering 
	\label{fig4}
\end{figure*}

  \begin{figure*}[tbp]
	\centering	
    \includegraphics[width=0.9\textwidth]{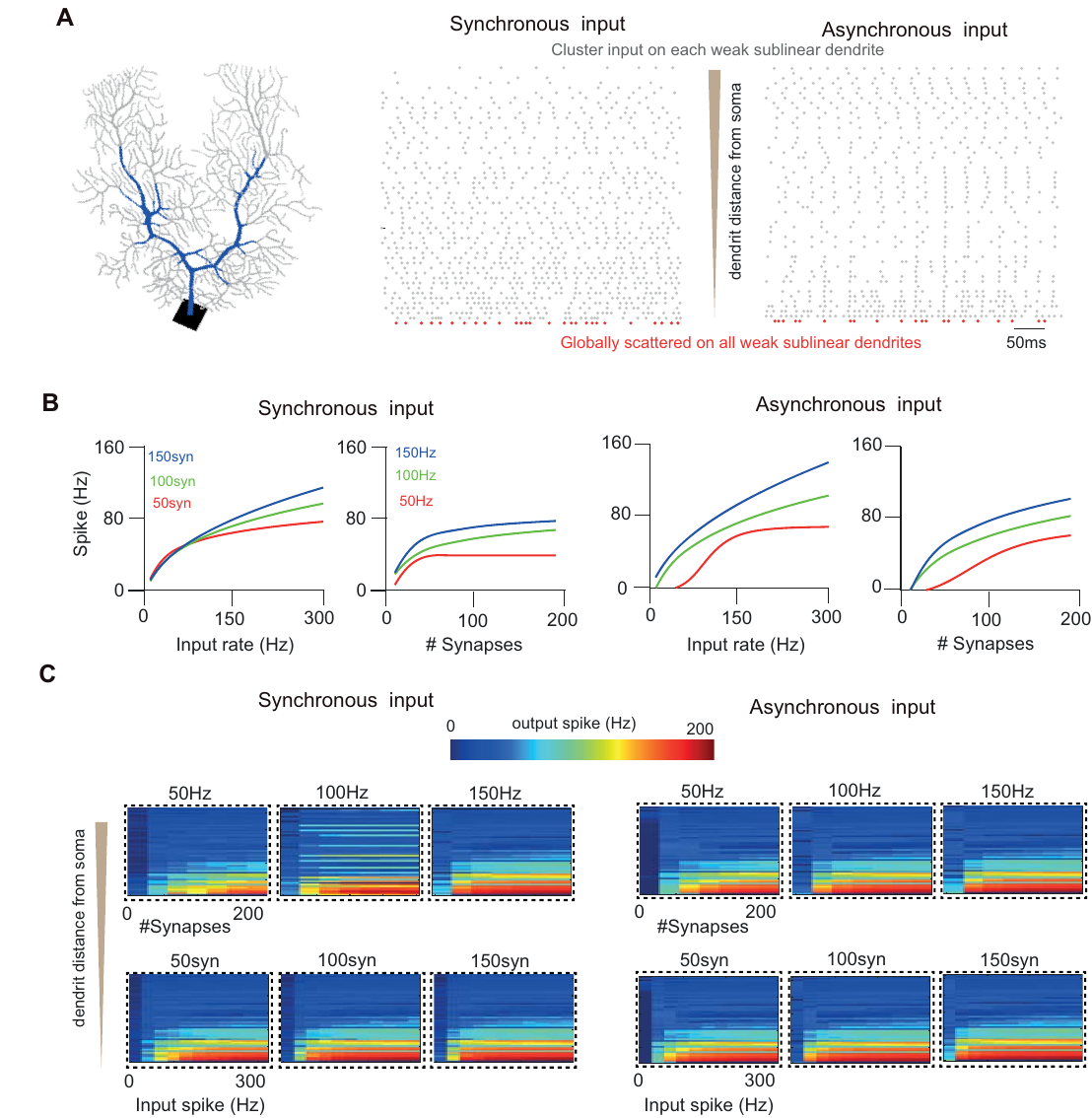}
	\caption{ Somatic spiking response is minimally affected by spatial and temporal synaptic input patterns on weak sublinear dendrites. 
    (A) (Left) main dendrites (blue) are defined by weak sublinearity less than 20\%. (Right) Raster plot of somatic spiking activity triggered by 50 synaptic synchronous and asynchronous inputs on main dendrites at 100Hz with the simulation of cluster input (gray) in each single dendrite and globally scattered (red) across all main dendrites. Each dot is a spike time. Each row in gray shows spiking in response to a single dendrite with decreasing order of dendritic distance to soma.
    (B)  Spiking activity vs. input rate using (left) synchronous and (right) asynchronous stimulation with 50, 100, and 150 synapses or 50, 100, and 150 Hz using globally scattered synaptic inputs. 
    (C) Similar to (B) but using cluster synaptic inputs on each single dendrite.    
	}
	\centering
	\label{fig5}
\end{figure*}

Given that PCs typically show a significantly high firing rate and install a large mumbler of dendrites, they receive a large number of input from granule cells over the entire
dendrite field. We then study how different spatial types of synaptic input can affect somatic firing through dendritic sublinearity. We selected dendrites with a sublinearity index greater than 20\% as spiny dendrites and grouped them into 24 branches along the main dendrites (Figure~\ref{fig3}A).  Characteristics, the length, diameter and area, of 24 branches are consistent with the methodology used in~\cite{zang2021cellular}, indicating significant variability among the branches and highlighting the structural diversity within the selected branches (Figure~\ref{suppfig5}). For each branch, only one dendritic segment connects it to the parent main dendrite. To simulate clustered synaptic input, 50 synapses were distributed on a single dendrite or branch and activated synchronously. Figure~\ref{fig3}B shows the somatic response by stimulating single dendrites (gray) and individual branches (colored) with a clustered input. Overall, the maximum membrane potential is greater when synapses are randomly distributed throughout the dendrites within a branch (scattered input) than when synapses are clustered on a single dendrite (cluster input), except for those dendrites close to the main trunk. Distance-dependent properties make proximal dendrites more responsive. 

Strong dendritic sublinearity makes the somatic spiking firing threshold more difficult to reach (Figure~\ref{fig3}C). The type of dendritic operation strongly contributes to the
nature of the resultant neuronal computation. For a neuron with sublinear dendrites, clustered synaptic activity will be less efficient at triggering a spike than scattered input distributed in different dendrites due to strong sublinearity (Figure~\ref{fig3}B). By varying the number of synaptic inputs, the input-output relationship obtained from single dendrites and the entire branch shows that the summation over multiple branches using scattered inputs can enhance the firing significantly (Figure~\ref{fig3}D). Using branch 1 (br 1) within the left main trunk, as an example, the input-output curves of all dendrites in branch 1 in Figure~\ref{fig3}D (left) show that the synapses are randomly distributed throughout the branch to produce a larger EPSP, except for dendrite 419 (top gray curve). Coactivation of branch1 with other branches can produce higher peak EPSP (Figure\ref{fig3}D, middle and right). 

To further examine how dendritic sublinear integration regulates somatic output under different spatial distributions, we considered three input strategies of synapses(Figure~\ref{fig4}A, top), using dendritic 419 in branch 1, which has the highest peak of EPSP, as an example. When all 50 synapses are activated synchronously, spikes are induced in all distribution strategies, but locally scattered has fewer spikes (Figure~\ref{fig4}A, middle). However, when synapses are activated asynchronously as the Poisson process, spikes are generated only under the global scattered distribution (Figure~\ref{fig4}A, bottom). We further increased the number of synapses and stimulation frequency and found that only the global scattered distribution induced spiking activity under asynchronous activation in branch 1 (Figure~\ref{fig4}B). In contrast, both cluster input and locally scattered input can not effectively produce spiking. Additionally, similar results are observed in the guinea pig (NMO\_00610) PC cell model (Figure~\ref{suppfig6}A). Specifically, only the globally scattered input distribution was capable of generating spikes, while both cluster input and locally scattered input failed to produce spiking activity, even at higher input rates.
However, changing the number of synapses and stimulation frequency during synchronous activation induces spike production under all distribution strategies(Figure~\ref{fig4}C).

The effect of sublinearity can be revealed using trunk dendrites with less than 20\% sublinearity index (Figure~\ref{fig5}A). We found that stimulation on these dendrites is more likely to induce spiking (Figure~\ref{fig5}A), independent of synaptic temporal activation mode, synchronous or asynchronous input, as well as spatial distribution strategy, scattered (Figure~\ref{fig5}B) or cluster (Figure~\ref{fig5}C) input. While the closer the synaptic distribution location is to soma, the more spikes are induced. These results are consistent with the notation that climbing fibers targeting trunk dendrites close to soma can trigger more complex spikes.

\subsection{Strong sublinearity suppress bursting spike}

  \begin{figure*}[tbp]
	\centering	
    \includegraphics[width=0.8\textwidth]{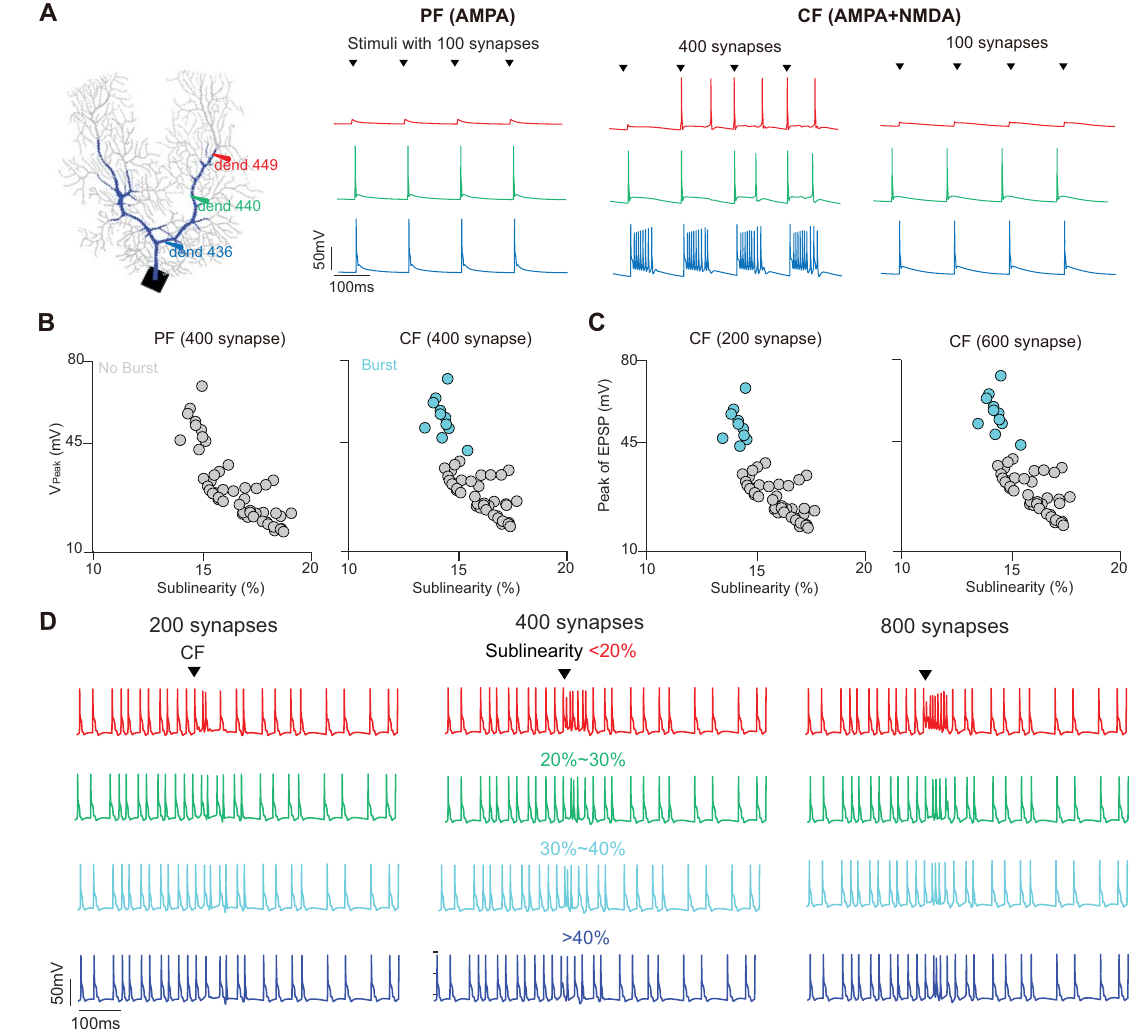}
	\caption{ Bursting complex spikes are suppressed by strong sublinearity. 
    (A) 400 synapses cluster input in three dendrites. Somatic voltage responses are triggered by each stimulation site with and without NMDARs. The triangle represents the time of stimulation.   
    (B) Scatter plots showing the peak of EPSP and sublinearity of main dendrites with NMDA and without NMDA. Cyan scatter and dendrites indicated that somatic has a burst response. 
    (C) Similar to B, but with 200 synapses (left) and 600 synapses (right) cluster input.
    (D) Strong sublinearity suppresses burst. PC installed with 1000 PF AMPA synapses stimulated at 30 Hz Poisson spikes and a CF with 200, 400, and 800 AMPA+NMDA synapses distributed on different sublinear dendrites.
	} 
	\centering
	\label{fig6}
\end{figure*}

To examine the view that climbing fiber (CF) input induces complex spikes (CSs) in PCs, we test the potential for CSs at different dendritic locations. Figure~\ref{fig6}A displays an example of cluster input synapses distributed at three sites. CSs as spiking bursts can not be induced by 100 synapses and can not be triggered with AMPA-like parallel fibers (PF). With 400 synapses, CSs occur in the presence of CFs with additional NMDA receptors (Figure~\ref{fig6}B). Dendrites with weaker sublinearity and larger peaks are more likely to produce CS when CF inputs, which is relatively independent of the number of synapses ( Figure~\ref{fig6}C). 

Purkinje cells can be categorized into two types, Zebrin positive (Z+) and Zebrin negative (Z-)~\cite{fernandez2024purkinje}. Z+ PCs are characterized by lower background firing rates and weaker responses to synaptic input, while Z- PCs exhibit higher background activity and stronger input responses~\cite{cheron2022electrophysiological}. We combined PF and CF inputs to simulate a more realistic scenario. Background excitatory input was mimicked by distributing PF synapses evenly onto spiny dendrites at 30 Hz, and CF inputs were mimicked by distributing synapses onto the different sublinear regions with different numbers of synapses (Figure~\ref{fig6}D). With 200 synapses, complex spikes occur when synapses are distributed in dendritic regions with a sublinearity of less than 20\%, causing increased somatic maximum spike rate. With 400 synapses, CS can not be induced when synapses are distributed only in dendrites with sublinearity greater than 40\%. When the input increases to 800 synapses, CSs occur more robustly and therefore the somatic maximum spike rate increases further. The similar results are observed in the guinea pig (NMO\_00610) PC (Figure~\ref{suppfig6}B). Overall, stronger sublinearity inhibits CS production. Under high-frequency (100 Hz) background stimulation, the occurrence of CSs is suppressed compared to low-frequency stimulation (Figure~\ref{suppfig7}). Strong nonlinear responses continue to suppress the generation of CSs, regardless of the background frequency. These results reveal the mechanism of bursting complex spikes highly depends on the dendritic sublinearity.

\subsection{Dendritic sublinearity implements the feature binding}

\begin{figure*}[tbp]
	\centering	
    \includegraphics[width=0.9\textwidth]{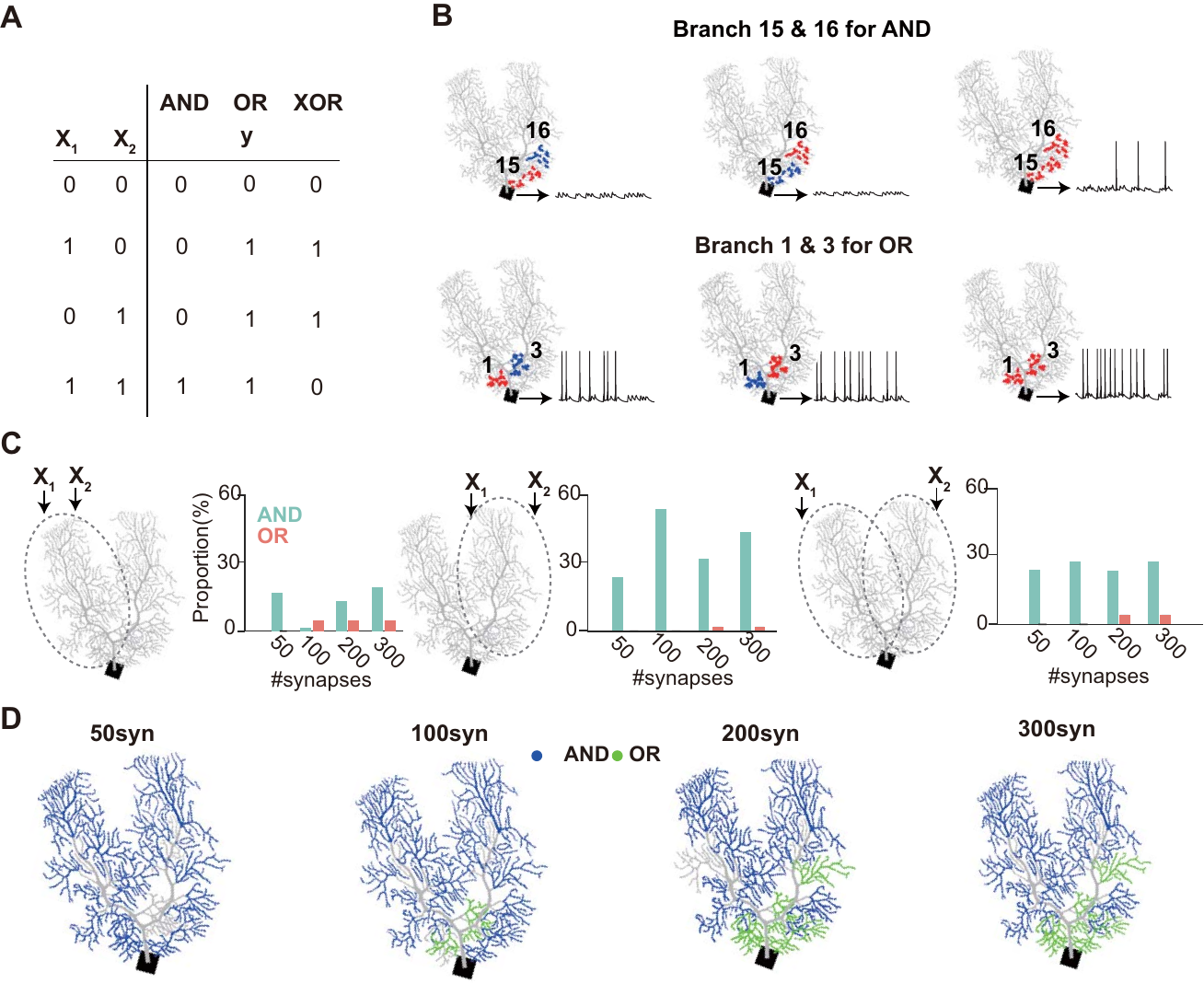} 
	\caption{Sublinear dendrites implement Boolean functions.
   (A) 24 branches with each representing a computing unit. Synapses are binary as 0 or 1 for the truth table of AND, OR, and XOR with two inputs ($ X_1$ and $ X_2$).  
   (B) Branches 15 \& 16 for AND, and branches 1 \& 3 for OR. Colored dots indicate the distribution location of synapses, blue for inactivation, and red for activation. 100 synapses with 50 Hz synchronous input on each branch. 
   (C)  Two synchronous inputs on the left, right, and both the main branches. The bar chart shows the ratio of pairs of branches for AND and OR. 
   (D) The distribution of branches participating in AND (blue) and OR (green) with different synapse inputs. Gray areas indicate regions not participating in any Boolean operations. 
	} 
	\centering
	\label{fig7}
\end{figure*}

  \begin{figure*}[tbp]
	\centering	
    \includegraphics[width=0.8\textwidth]{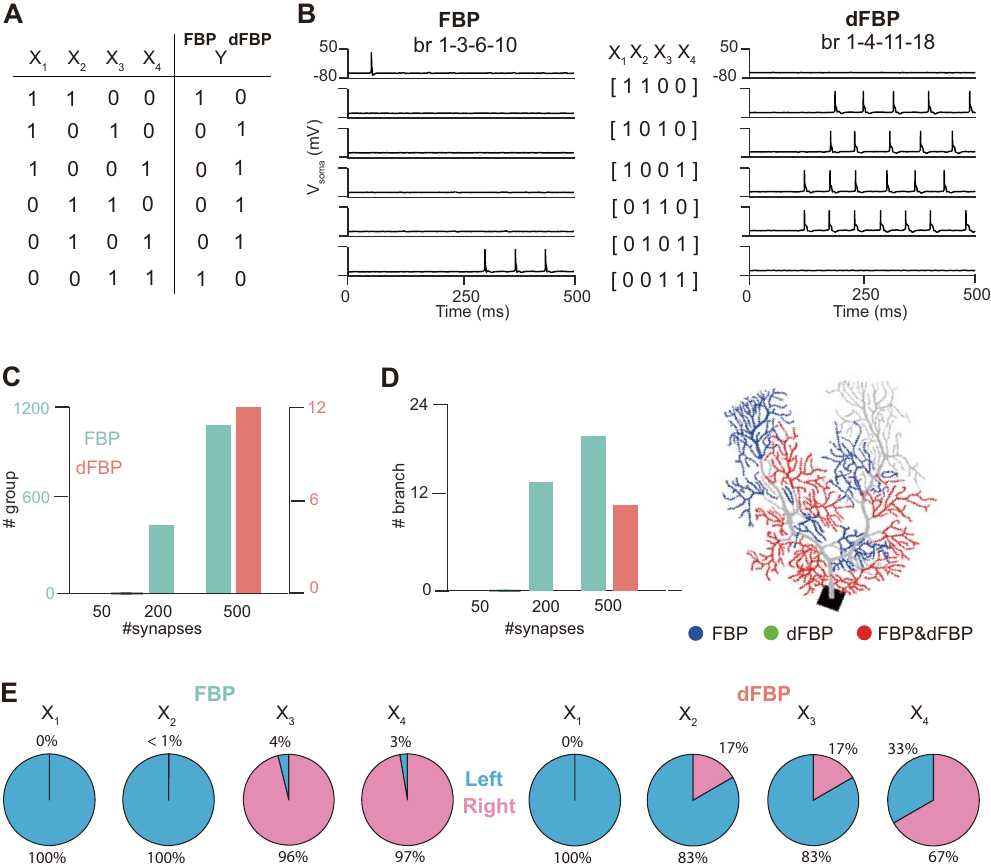}  
	\caption{Sublinear dendrites implement the feature binding problem (FBP) and dual feature binding problem (dFBP). 
    (A) FBP and dFBP with 4 variables.
    (B) Examples of FBP and dFBP computed by a group of 4 branches. The somatic voltage trace in response to different input vectors where one active input variable is represented by an asynchronous input with 500 synapses at 50 Hz.
   (C) The number of branch ensembles (groups) for FBP and dFBP with an input of different numbers of synapses.
   (D) The number of branches for FBP and dFBP and their distribution on the dendritic tree. Blue regions indicate participation only in feature binding problem (FBP), green regions indicate participation only in double feature binding problem (dFBP), and red regions indicate participation in both FBP and dFBP. Gray areas indicate regions not participating in any feature binding operations. 
   (E) The distribution of 4 input vectors on the left and right main branches when implementing FBP and dFBP.   
	} 
	\centering
	\label{fig8}
\end{figure*}

We finally examine the functional role of dendritic sublinearity using classical computing tasks. Our PC model has 24 branches based on dendritic sublinearity, where each branch can be a single computing unit. We first consider the computing task of the Boolean function. For this, synapses are assigned with a binary value of 0 or 1. The final output is 1 if the PC fire spikes, otherwise 0 for no spike. In this context the input-output relationship can be described by a unique truth table, corresponding to a Boolean function (Figure~\ref{fig7}A). The truth table describes three simple Boolean functions: OR, AND, and XOR. To compute this, we applied synaptic inputs on a pair of branches, went through all the pairs out of 24 branches, and recorded the somatic membrane response (Figure~\ref{fig7}B). Examples show that the neural model can implement either the AND (branches 15 and 16), or OR (branches 1 and 3) Boolean function. However, no example of XOR was found. The particular PC has two major trunk dendrites, so we exhausted all possible pairs on both trunks and counted the percentage of pairs for each Boolean function (Figure~\ref{fig7}C). In the left main trunk, only AND can be performed when the number of synaptic inputs is 50, and the cell starts to perform OR as the number of synapses increases. At the right main trunk, the PC implements a higher proportion of ANDs and requires a higher number of synapses to perform OR. When the two branches are from the two main trunks, the change in the number of synapses has little effect on performing AND, while more difficult to implement OR. In general, the cell prefers to perform AND, and more synaptic inputs are required to perform OR with synchronous inputs (Figure~\ref{fig7}D). Such observations are relatively independent of the temporal input protocols and preserved when asynchronous inputs were used (Figure~\ref{suppfig8}).

 \begin{figure*}[tbp] 
	\centering	
    \includegraphics[width=1\textwidth]{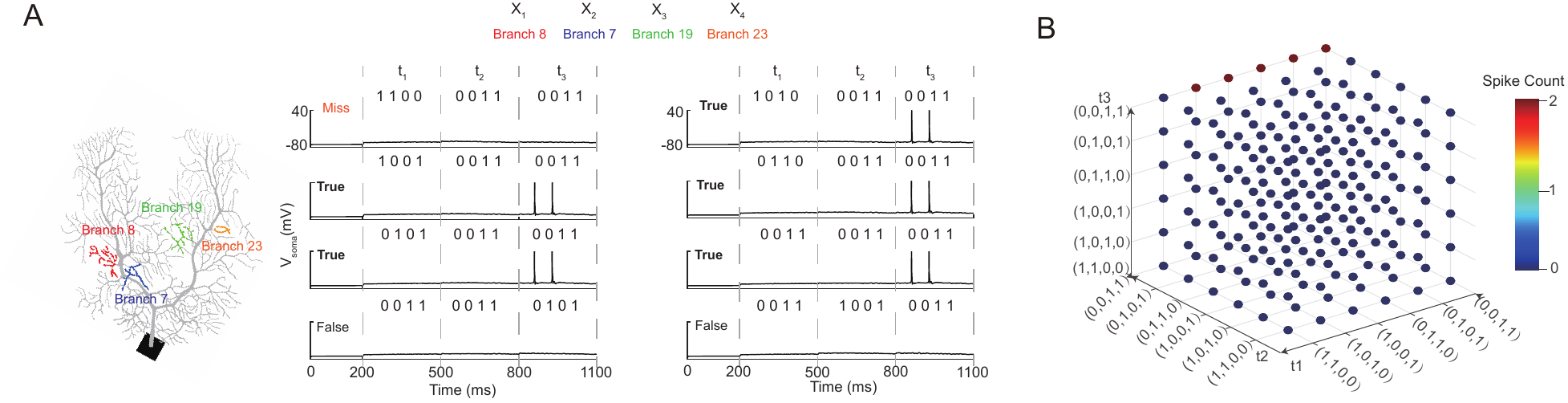}
	\caption{  Sublinear dendrites can achieve a time-dependent feature binding.
    (A) The diagram illustrates three time steps ($t_1$, $t_2$, $t_3$) with four features distributed across dendritic branches 8, 7, 19, and 23. The feature vectors ($ X_1$, $ X_2$, $ X_3$, $ X_4$) represent activation states with 0 for inactive and 1 for active. FBP is successful if spikes occur only when the features ($X_3$, $ X_4$) are activated during the time steps ($t_2$, $t_3$), indicating by True, otherwise False. Miss indicates the failure of FBP.  
    %In the last row of the figure, altering the activation sequence of the features fails to produce spikes, demonstrating the order-dependent of this feature binding process.
    (B) Spike count across all 216 possible combinations of feature activations, where the X, Y, and Z axes represent feature vectors from three time periods. The color scale indicates the spike count, with 5 True cases and only 1 Miss case, as shown in (A).
        }
	\centering
	\label{fig9}
\end{figure*}

We then consider a more difficult computing task of feature binding problem (FBP)~\cite{tran2015contribution}. For the input, we considered that the four input variables $ X_i, i=1,2,3,4$ corresponded to four afferent neural ensembles of 500 neurons each in each branch, and that an ensemble was active, signaling $ X_i=1$. For the output, we considered a single somatic spike as the response to an input vector, such that if the neuron fires a spike then $ Y=1$, and no spike as $ Y=0$. Thus, to implement the FBP successfully, the cell should only fire spike in response to the input vectors $ X=[1,1,0,0]$ and $[0,0,1,1]$, and not to spike for the remaining four input vectors (Figure~\ref{fig8}A). The dual feature binding problem (dFBP) is the opposite. The neuron should not fire spikes in response to the input vectors $ X=[1,1,0,0]$ and $ [0,0,1,1]$, and only spike for the remaining four input vectors. Indeed, we found the neuron can implement FBP, for example, with a combination of branches 1, 3, 6, and 10 (Figure~\ref{fig8}B, left), and dFBP with a combination of branches 1, 4, 11, and 18 (Figure~\ref{fig8}B, right). Next, we want over 24 branches with a group of four branches for all input vectors and recorded the number of groups of branches for both FBP and dFBP tasks (Figure~\ref{fig8}C) with asynchronous inputs. By varying the number of synaptic inputs, we found that implementing dFBP requires more inputs while FBP requires an appropriate number of synapses. As the number of synapses increases, the number of branches involved in the implementation of FBP increases. Compared to FBP, there are significantly fewer groups that can implement dFBP. Particularly, nearly all branches can be involved in FBP, while only half of the branches for dFBP over all 24 branches (Figure~\ref{fig8}D). We analyzed how the branches involved in FBP and dFBP are distributed over the four input vectors, and found that $ X_1$ and $ X_2$ tend to sit in the one trunk and $ X_3$ and $ X_4$ on the other trunk (Figure~\ref{fig8}E), due to that the two branches produce stronger sublinearity when they are on different main dendrites. For instance, the observed EPSP $ O(EPSP_{br1+br2})$ will be smaller than the expected $ E(EPSP_{br1+br2})$ when two branches on different main dendrites. Generally speaking, there is further sublinearity between different branch combinations. Such observations highly depend on the temporal input protocol. When synchronous input was used, dFBP was more easily implemented with fewer numbers of synaptic inputs (Figure~\ref{suppfig9}). In addition, the sublinearity index of each branch is calculated by the average of the sublinearity index of all dendrites within the branch (Figure~\ref{suppfig10}). It was observed that when inputs $  X_1$  and $  X_2$ are both distributed in branches with weak sublinearity, the neuron struggles to perform AND operations (Figure~\ref{suppfig11}A). However, the likelihood of achieving AND operations increases when at least one of the inputs is distributed in a branch with moderate sublinearity. Additionally, achieving effective feature binding problem (FBP) and double feature binding problem (dFBP) requires a balance of sublinearity within the four branches, emphasizing the need for a moderate balance of sublinearity to support these complex computations (Figure~\ref{suppfig11}B).

Moreover, nonlinear dendrites can achieve feature binding with time-dependent signals (Figure~\ref{fig9}). The feature binding process is shown across three time steps ($t_1, t_2, t_3$), with four features distributed across dendritic branches (8, 7, 19, and 23). The membrane potential plots illustrate that spikes occur only when the third ($X_3$) and fourth ($X_4$) features are activated during the second ($t_2$) and third ($t_3$) time steps (Figure~\ref{fig9}A). Importantly, changing the activation sequence fails to produce spikes, highlighting the order-dependent nature of this feature binding process. In this particular example, there are five True cases where FBP is successful, while only one Miss case shows the failure of FBP without spikes. This observation is seen from spike count across all 216 possible combinations of feature activations (Figure~\ref{fig9}B). These results confirm that the dendritic nonlinearity allows precise temporal integration, enabling the neuron to selectively bind spatial and temporal features.

\section{ Discussion}

In the realm of single-neuron computation, the rational underpinning lies in the intricacies of nonlinear dendritic behavior. Here utilizing a Purkinje cell model with the detailed dendritic morphology, we revealed a distinctive feature of sublinear synaptic integration at different parts of dendrites. It is this inherent sublinearity within the dendritic architecture that empowers cerebellar PCs to implement binding of different features, potentially sending from parallel and climbing fibers. Our study sheds light on the computational prowess of PCs, emphasizing the pivotal role played by dendritic sublinearity in facilitating the realization of sophisticated neural computations.

\subsection{Sublinear dendritic integration in Purkinje cells}

Neurons within the cerebellum manifest complex nonlinear properties, a phenomenon deemed critical in orchestrating network functionality~\cite{d2018physiology}. The dendrites of cerebellar interneurons precisely filter synaptic response times and amplitudes, engendering sublinear input-output relationships. Elucidating the transformation of temporally and spatially distributed inputs from granule cells constitutes a crucial pursuit in understanding their operational dynamics~\cite{abrahamsson2012thin,tran2016differential}. Notably, the morphological attributes of neurons introduce a consequential dimension, wherein the amplitude and configuration of local EPSPs are subject to the nuanced intricacies of neuronal form~\cite{magee2000dendritic}.

In contrast, PCs present a more complex dendritic morphology, necessitating the integration of a greater multitude of synaptic inputs. Our investigation reveals a distinctive property: PC dendrites manifest sublinear synaptic integration, with the degree of sublinearity accentuated in thin dendrites as opposed to their thicker counterparts. Remarkably, this position-dependent nonlinearity in synaptic integration is reminiscent of observations in CA1 pyramidal neurons~\cite{cash1999linear}. A plausible rationale lies in the elevated input resistance of thin dendrites, mitigating the driving forces for synaptic current flow concerning their thicker counterparts. This underlines the PC's ability to differentiate similar inputs based on the synaptic distribution position. Moreover, the integration of inputs from granule cells, strategically distributed along the spiny dendrites, undergoes effective filtering to preclude saturation.

Intriguingly, PC dendrites exhibit sensitivity to synaptic inputs at the millisecond scale. This contrasts with pyramidal cell dendrites, which, within short temporal intervals ($0-5$ ms), exhibit either linearity or slight supralinearity~\cite{schiller2000nmda, urban1998active}. However, as temporal intervals extend ($5-100$ ms), EPSP summation in pyramidal cell dendrites becomes sublinear~\cite{magee1999dendritic, margulis1998temporal}. The heightened sensitivity of PC dendrites to millisecond-scale synaptic information potentially accounts for the precise synchronous firing of simple spikes in on-beam Purkinje cells during cerebellar behaviors~\cite{de2008high, heck2007beam, person2012purkinje}. In addition, Purkinje cells control the coordination and adaptation of sensorimotor behavior during visuomotor behavior by nonlinearly integrating signals~\cite{knogler2019motor}. Nonlinearity allows Purkinje cells to respond sensitively to minute changes in input signals while maintaining a stable response across a broad range of input signal intensities. This means that Purkinje cells can precisely regulate the inhibitory signals they send to the deep nuclei of the cerebellum, thus finely controlling movement~\cite{heiney2014precise}. In concert, the observed sublinear integration in PCs assumes the role of an activity-dependent spatial filter, contributing substantively to the adaptive filter behavior within the cerebellar cortex. 

\subsection{Dendrites sublinearity affects neuronal input-output dynamics}

The intricate interplay between theory and experimentation has consistently stressed the critical role played by dendrites in shaping the input-output neuronal dynamics, conferring formidable computing capabilities upon individual neurons~\cite{larkum2009synaptic, branco2010dendritic, major2013active, lafourcade2022differential}. The influence of nonlinear dendritic integration in actively modulating neuronal responses has been experimentally demonstrated across diverse cortical areas~\cite{grienberger2014nmda, smith2013dendritic, palmer2014nmda}. Theoretical insights suggest that dendritic superlinearity makes neurons sensitive to clustering inputs, while sublinearity enables sensitivity to scattering inputs~\cite{caze2013passive,tran2015contribution}. 
 
Remarkably, in the somatosensory cortex, features are localized within single dendrites~\cite{takahashi2012locally}, while synapses in the visual and auditory cortex's layer 2/3 pyramidal neurons exhibit maximal responsiveness to inputs carrying distinct sensory features~\cite{jia2010dendritic, chen2011functional}. Consequently, this global scatter strategy aligns with the neuron's responsiveness to multi-feature discrete distributions. Furthermore, our investigation illuminates that weak sublinearity in dendrites favors the generation of bursts. Notably, the principal dendrite establishing a synaptic connection between climbing fibers and PC emerges as a region of weak sublinearity. This finding posits that climbing fiber inputs can more readily induce PCs to generate complex spikes (CSs).
Simple spike (SS) activity and the wave-like pattern of CS activity have been shown to correlate with Zebrin expression~\cite{zhou2014cerebellar}. Previous studies have shown that in Zebrin positive (Z+) PCs, CSs exhibit a stronger modulatory effect on SSs compared to Z- PCs~\cite{tang2017heterogeneity}. Our findings reveal that high-frequency background stimulation suppresses CS generation, whereas low-frequency stimulation allows CS to occur more prominently. This is in line with the observation of the stronger CS modulation of SS in Z+ PCs shown in experiments~\cite{tang2017heterogeneity}, where the lower background activity favors CS occurrence under low-frequency conditions. We demonstrated that strong nonlinear dendritic responses consistently suppress CS generation, regardless of background frequency, offering fresh insights into the distinct computational roles of Z+ and Z- PCs in cerebellar processing.

\subsection{Dendritic nonlinearity implements feature binding }

Should dendrites merely serve as collectors of input, performing linear summation at the soma, a single neuron would be confined to computing solely linearly separable input-output functions. However, experimental evidence has revealed the capacity of neuronal dendrites to undergo nonlinear synaptic integration~\cite{koch1983nonlinear, xu2012nonlinear, polsky2004computational, poirazi2020illuminating}. Neurons exhibiting the nonlinear synaptic integration transcend the limitation, calculating both linearly separable and non-separable functions. Theoretical investigations underscore the transformative role of dendritic nonlinearity, empowering individual neurons to execute linearly inseparable functions~\cite{caze2012spiking, tran2015contribution, caze2013passive, caze2021all}, with recent experiments showcasing the capability of cerebellar stellate cells to solve feature binding problems~\cite{caze2023demonstration}.

In our study, we unveil the capability of PCs with sublinear dendritic characteristics, demonstrating their capacity to perform Boolean operations, primarily preferred AND operations. As the synaptic inputs increase, the execution of OR operations becomes more pronounced. Remarkably, the facilitation of AND operations is prominent when two inputs are strategically distributed on one of the two main trunk dendrites, however, the XOR operation is unachievable here, unlike human pyramidal cells~\cite{gidon2020dendritic}. Furthermore, the sublinear attributes of PC dendrites contribute to implementing feature binding problems, where both asynchronous and synchronous input can realize dual feature binding problems. Intriguingly, during the implementation of the feature binding problem, inputs appear equally distributed among the main primary dendrites. This observation aligns with the notion that more primary dendrites are requisite for complex functions. Notably, human PCs~\cite{masoli2023human,busch2023climbing}, characterized by more complex dendritic structures and enhanced computing capabilities compared to rodents, typically exhibit 2-3 main trunks, in contrast to most rodents with only a single main trunk. The presence of multiple dendritic trunks is correlated with climbing fiber connections and long-term synaptic plasticity~\cite{najafi2013beyond} and this innervation pattern can generate independent computational compartments within a single Purkinje cell~\cite{busch2023climbing}. Our findings further propose that branches between main dendrites boost sublinear synaptic integration, thereby augmenting the computational capability of a single neuron. Together with other recent studies, neurons endowed with nonlinear dendritic characteristics surpass their linear counterparts in computing more complex tasks~\cite{kaifosh2016mnemonic}.

\subsection{General functional implication of PC dendritic computation}

The computational capabilities of PCs are fundamentally shaped by their complex dendritic properties, which underlie their various functional functions in the cerebellum. Studies have shown that dendritic morphology and synaptic plasticity enable PCs to perform pattern recognition and temporal sequence encoding~\cite{steuber2007cerebellar,majoral2020model,zang2018voltage}. The distribution of ion channels along dendrites facilitates multiplexed coding, allowing PCs to process multiple input streams simultaneously~\cite{zang2021cellular}. 
Interestingly, a similar demand for efficient multi-input integration appears in artificial systems. In computer vision, the ability to accurately process multiple input streams under affine transformations is essential for robust pose estimation and motion analysis~\cite{guan2020minimal, guan2021relative}, reflecting a shared computational principle between biological and artificial domains.
Nonlinear integration of dendritics further improves temporal discrimination, allowing advanced computations~\cite{tamura2023discrimination,xu2012nonlinear}. Furthermore, increased dendritic complexity, as observed in human PCs, is associated with superior computational performance~\cite{masoli2024human}. Dendritic nonlinearities enhance the computational capacity of neurons by facilitating multi-layered processing of information within individual cells.

Here we demonstrated a spatial gradient in sublinearity, with stronger suppression in thinner dendrites, which can improve PC computational capabilities by facilitating selective input filtering, localized dendritic processing, and efficient spatiotemporal integration across dendritic regions, as demonstrated in cerebellar interneurons~\cite{abrahamsson2012thin}. Theoretical work by~\cite{tran2015contribution} proposed that sublinear dendritic integration enables scatter sensitivity, favoring distributed inputs. Interestingly, similar behavior has also been observed experimentally in cerebellar stellate cells, as reported in~\cite{caze2024demonstration}. Our study confirms this in Purkinje cells, showing that dendritic sublinearity enforces a global scatter strategy, where only globally distributed inputs induce spiking. 
This outcome suggests a resilience of PCs against firing induced by clustered granule cell inputs, attributed to the abundance and sparsity of granule cells, thus avoiding the triggering of PC firing by a limited subset of granule cells. This selective filtering mechanism shares conceptual similarity with strategies developed in robust estimation problems in computer vision, where global consistency is prioritized over local agreement to resist noise and outliers~\cite{guan2022affine}.

Moreover, the FBP is particularly intriguing because it allows neurons to integrate multiple feature combinations into a cohesive representation~\cite{yu2023binding,von1999and}. Analogously, in computer vision, binding multiple local features under geometric constraints is critical for achieving reliable correspondence estimation~\cite{sun2025learning}, a problem closely related to integrating distributed and heterogeneous inputs, as observed in biological neurons.
In our study, by integrating temporal information into the feature binding problem, we demonstrated that PCs can selectively bind spatial and temporal features. This highlights the role of PCs in addressing complex problems that involve distributed and temporally ordered inputs~\cite{ito2008control,ujfalussy2018global}. As described in~\cite{masoli2024human}, the increased dendritic complexity and computational capacity of human Purkinje cells support our findings, suggesting that the dendritic nonlinearity and spatiotemporal integration observed in rodents PCs could function even more efficiently in the more complex dendritic architecture of human PCs.
%Moreover, \cite{li2021deep,allam2019achieving} utilize the same Purkinje cell (NMO-00865) model to explore principles of efficiency and adaptability in biological and artificial systems. \cite{li2021deep} demonstrates how distributed input strategies optimize resource distribution, aligning with our findings that the global scatter strategy enhances dendritic computation in PCs. \cite{allam2019achieving} focuses on neuroplasticity-driven adaptability, paralleling our results showing spatiotemporal feature binding facilitated by dendritic nonlinearity. Together, these studies align with our findings by highlighting how dendritic sublinearity optimizes distributed input integration and supports high-dimensional computations. 

PCs also receive inhibitory input from interneurons that can significantly impact PC dynamics~\cite{tang2021modulation}, which is essential for the precise timing and coordination of muscular activities controlled by the cerebellum~\cite{pouzat1997developmental, dizon2011role}. %The effect of inhibitory input on the nonlinear synaptic integration ability of Purkinje cells has not been studied in depth here. However, it is obvious that 
Inhibition could regulate the nonlinear synaptic integration ability of PCs. Such regulation may also depend on the input location of specific dendrites. Future research would investigate the impact of inhibitory input on dendritic nonlinearity to fully understand the complex mechanisms underpinning the PC function. With the recent view on the diversity of the functional role of the cerebellum~\cite{dezeeuwDiversityDynamismCerebellum2021} and the interaction between cerebellum and neocortex~\cite{zhuActivityMapCorticocerebellar2023}, the nonlinear dendritic integration of Purkinje cells could potentially contribute to more complex cortical dynamics.

\section*{Acknowledgments}
This work was partially supported by the National Natural Science Foundation of China under Grant No. 62302016, 62422601, 62176003, and 62088102, and the UK Royal Society Newton Advanced Fellowship Grant No. NAF-R1-191082

\section*{Data availability}
The code is available at https://github.com/ydtang/Nonlinearity-model.

\bibliographystyle{IEEEtran}
\bibliography{main}

% Generated by IEEEtran.bst, version: 1.14 (2015/08/26)
\begin{thebibliography}{10}
\providecommand{\url}[1]{#1}
\csname url@samestyle\endcsname
\providecommand{\newblock}{\relax}
\providecommand{\bibinfo}[2]{#2}
\providecommand{\BIBentrySTDinterwordspacing}{\spaceskip=0pt\relax}
\providecommand{\BIBentryALTinterwordstretchfactor}{4}
\providecommand{\BIBentryALTinterwordspacing}{\spaceskip=\fontdimen2\font plus
\BIBentryALTinterwordstretchfactor\fontdimen3\font minus \fontdimen4\font\relax}
\providecommand{\BIBforeignlanguage}[2]{{%
\expandafter\ifx\csname l@#1\endcsname\relax
\typeout{** WARNING: IEEEtran.bst: No hyphenation pattern has been}%
\typeout{** loaded for the language `#1'. Using the pattern for}%
\typeout{** the default language instead.}%
\else
\language=\csname l@#1\endcsname
\fi
#2}}
\providecommand{\BIBdecl}{\relax}
\BIBdecl

\bibitem{davie2006dendritic}
J.~T. Davie, M.~H. Kole, J.~J. Letzkus, E.~A. Rancz, N.~Spruston, G.~J. Stuart, and M.~H{\"a}usser, ``Dendritic patch-clamp recording,'' \emph{Nature protocols}, vol.~1, no.~3, pp. 1235--1247, 2006.

\bibitem{spruston2008pyramidal}
N.~Spruston, ``Pyramidal neurons: dendritic structure and synaptic integration,'' \emph{Nature Reviews Neuroscience}, vol.~9, no.~3, pp. 206--221, 2008.

\bibitem{spruston2016principles}
N.~Spruston, G.~Stuart, and M.~H{\"a}usser, ``Principles of dendritic integration,'' \emph{Dendrites}, vol. 351, no. 597, p.~1, 2016.

\bibitem{lafourcade2022differential}
M.~Lafourcade, M.-S.~H. van~der Goes, D.~Vardalaki, N.~J. Brown, J.~Voigts, D.~H. Yun, M.~E. Kim, T.~Ku, and M.~T. Harnett, ``Differential dendritic integration of long-range inputs in association cortex via subcellular changes in synaptic {AMPA}-to-{NMDA} receptor ratio,'' \emph{Neuron}, vol. 110, no.~9, pp. 1532--1546, 2022.

\bibitem{branco2010single}
T.~Branco and M.~H{\"a}usser, ``The single dendritic branch as a fundamental functional unit in the nervous system,'' \emph{Current opinion in neurobiology}, vol.~20, no.~4, pp. 494--502, 2010.

\bibitem{london2005dendritic}
M.~London and M.~H{\"a}usser, ``Dendritic computation,'' \emph{Annu. Rev. Neurosci.}, vol.~28, pp. 503--532, 2005.

\bibitem{koch2000role}
C.~Koch and I.~Segev, ``The role of single neurons in information processing,'' \emph{Nature neuroscience}, vol.~3, no.~11, pp. 1171--1177, 2000.

\bibitem{poirazi2020illuminating}
P.~Poirazi and A.~Papoutsi, ``Illuminating dendritic function with computational models,'' \emph{Nature Reviews Neuroscience}, vol.~21, no.~6, pp. 303--321, 2020.

\bibitem{stuart2015dendritic}
G.~J. Stuart and N.~Spruston, ``Dendritic integration: 60 years of progress,'' \emph{Nature neuroscience}, vol.~18, no.~12, pp. 1713--1721, 2015.

\bibitem{makarov2023dendrites}
R.~Makarov, M.~Pagkalos, and P.~Poirazi, ``Dendrites and efficiency: Optimizing performance and resource utilization,'' \emph{Current Opinion in Neurobiology}, vol.~83, p. 102812, 2023.

\bibitem{cash1999linear}
S.~Cash and R.~Yuste, ``Linear summation of excitatory inputs by {CA1} pyramidal neurons,'' \emph{Neuron}, vol.~22, no.~2, pp. 383--394, 1999.

\bibitem{polsky2004computational}
A.~Polsky, B.~W. Mel, and J.~Schiller, ``Computational subunits in thin dendrites of pyramidal cells,'' \emph{Nature neuroscience}, vol.~7, no.~6, pp. 621--627, 2004.

\bibitem{larkum1999new}
M.~E. Larkum, J.~J. Zhu, and B.~Sakmann, ``A new cellular mechanism for coupling inputs arriving at different cortical layers,'' \emph{Nature}, vol. 398, no. 6725, pp. 338--341, 1999.

\bibitem{branco2010dendritic}
T.~Branco, B.~A. Clark, and M.~H{\"a}usser, ``Dendritic discrimination of temporal input sequences in cortical neurons,'' \emph{Science}, vol. 329, no. 5999, pp. 1671--1675, 2010.

\bibitem{major2013active}
G.~Major, M.~E. Larkum, and J.~Schiller, ``Active properties of neocortical pyramidal neuron dendrites,'' \emph{Annual review of neuroscience}, vol.~36, pp. 1--24, 2013.

\bibitem{branco2011synaptic}
T.~Branco and M.~H{\"a}usser, ``Synaptic integration gradients in single cortical pyramidal cell dendrites,'' \emph{Neuron}, vol.~69, no.~5, pp. 885--892, 2011.

\bibitem{losonczy2006integrative}
A.~Losonczy and J.~C. Magee, ``Integrative properties of radial oblique dendrites in hippocampal {CA1} pyramidal neurons,'' \emph{Neuron}, vol.~50, no.~2, pp. 291--307, 2006.

\bibitem{gasparini2006state}
S.~Gasparini and J.~C. Magee, ``State-dependent dendritic computation in hippocampal {CA1} pyramidal neurons,'' \emph{Journal of Neuroscience}, vol.~26, no.~7, pp. 2088--2100, 2006.

\bibitem{hu2010dendritic}
H.~Hu, M.~Martina, and P.~Jonas, ``Dendritic mechanisms underlying rapid synaptic activation of fast-spiking hippocampal interneurons,'' \emph{Science}, vol. 327, no. 5961, pp. 52--58, 2010.

\bibitem{carter2007timing}
A.~G. Carter, G.~J. Soler-Llavina, and B.~L. Sabatini, ``Timing and location of synaptic inputs determine modes of subthreshold integration in striatal medium spiny neurons,'' \emph{Journal of Neuroscience}, vol.~27, no.~33, pp. 8967--8977, 2007.

\bibitem{abrahamsson2012thin}
T.~Abrahamsson, L.~Cathala, K.~Matsui, R.~Shigemoto, and D.~A. DiGregorio, ``Thin dendrites of cerebellar interneurons confer sublinear synaptic integration and a gradient of short-term plasticity,'' \emph{Neuron}, vol.~73, no.~6, pp. 1159--1172, 2012.

\bibitem{tran2016differential}
A.~Tran-Van-Minh, T.~Abrahamsson, L.~Cathala, and D.~A. DiGregorio, ``Differential dendritic integration of synaptic potentials and calcium in cerebellar interneurons,'' \emph{Neuron}, vol.~91, no.~4, pp. 837--850, 2016.

\bibitem{xu2012nonlinear}
N.-l. Xu, M.~T. Harnett, S.~R. Williams, D.~Huber, D.~H. O’Connor, K.~Svoboda, and J.~C. Magee, ``Nonlinear dendritic integration of sensory and motor input during an active sensing task,'' \emph{Nature}, vol. 492, no. 7428, pp. 247--251, 2012.

\bibitem{wilson2016orientation}
D.~E. Wilson, D.~E. Whitney, B.~Scholl, and D.~Fitzpatrick, ``Orientation selectivity and the functional clustering of synaptic inputs in primary visual cortex,'' \emph{Nature neuroscience}, vol.~19, no.~8, pp. 1003--1009, 2016.

\bibitem{takahashi2016active}
N.~Takahashi, T.~G. Oertner, P.~Hegemann, and M.~E. Larkum, ``Active cortical dendrites modulate perception,'' \emph{Science}, vol. 354, no. 6319, pp. 1587--1590, 2016.

\bibitem{lavzin2012nonlinear}
M.~Lavzin, S.~Rapoport, A.~Polsky, L.~Garion, and J.~Schiller, ``Nonlinear dendritic processing determines angular tuning of barrel cortex neurons in vivo,'' \emph{Nature}, vol. 490, no. 7420, pp. 397--401, 2012.

\bibitem{kaifosh2016mnemonic}
P.~Kaifosh and A.~Losonczy, ``Mnemonic functions for nonlinear dendritic integration in hippocampal pyramidal circuits,'' \emph{Neuron}, vol.~90, no.~3, pp. 622--634, 2016.

\bibitem{tzilivaki2019challenging}
A.~Tzilivaki, G.~Kastellakis, and P.~Poirazi, ``Challenging the point neuron dogma: {FS} basket cells as 2-stage nonlinear integrators,'' \emph{Nature communications}, vol.~10, no.~1, p. 3664, 2019.

\bibitem{katz2009synapse}
Y.~Katz, V.~Menon, D.~A. Nicholson, Y.~Geinisman, W.~L. Kath, and N.~Spruston, ``Synapse distribution suggests a two-stage model of dendritic integration in {CA1} pyramidal neurons,'' \emph{Neuron}, vol.~63, no.~2, pp. 171--177, 2009.

\bibitem{koch1983nonlinear}
C.~Koch, T.~Poggio, and V.~Torre, ``Nonlinear interactions in a dendritic tree: localization, timing, and role in information processing.'' \emph{Proceedings of the National Academy of Sciences}, vol.~80, no.~9, pp. 2799--2802, 1983.

\bibitem{caze2013passive}
R.~D. Caz{\'e}, M.~Humphries, and B.~Gutkin, ``Passive dendrites enable single neurons to compute linearly non-separable functions,'' \emph{PLoS computational biology}, vol.~9, no.~2, p. e1002867, 2013.

\bibitem{caze2023demonstration}
R.~D. Caz{\'e}, A.~Tran-Van-Minh, B.~S. Gutkin, and D.~A. DiGregorio, ``Demonstration that sublinear dendrites enable linearly non-separable computations,'' \emph{bioRxiv}, pp. 2023--06, 2023.

\bibitem{poirazi2003pyramidal}
P.~Poirazi, T.~Brannon, and B.~W. Mel, ``Pyramidal neuron as two-layer neural network,'' \emph{Neuron}, vol.~37, no.~6, pp. 989--999, 2003.

\bibitem{beniaguev2021single}
D.~Beniaguev, I.~Segev, and M.~London, ``Single cortical neurons as deep artificial neural networks,'' \emph{Neuron}, vol. 109, no.~17, pp. 2727--2739, 2021.

\bibitem{li2020power}
X.~Li, J.~Tang, Q.~Zhang, B.~Gao, J.~J. Yang, S.~Song, W.~Wu, W.~Zhang, P.~Yao, N.~Deng \emph{et~al.}, ``Power-efficient neural network with artificial dendrites,'' \emph{Nature Nanotechnology}, vol.~15, no.~9, pp. 776--782, 2020.

\bibitem{chavlis2021drawing}
S.~Chavlis and P.~Poirazi, ``Drawing inspiration from biological dendrites to empower artificial neural networks,'' \emph{Current opinion in neurobiology}, vol.~70, pp. 1--10, 2021.

\bibitem{wu2023mitigating}
X.~Wu, P.~Zhao, Z.~Yu, L.~Ma, K.-W. Yip, H.~Tang, G.~Pan, and T.~Huang, ``Mitigating communication costs in neural networks: The role of dendritic nonlinearity,'' \emph{arXiv preprint arXiv:2306.11950}, 2023.

\bibitem{galakhova2022evolution}
A.~Galakhova, S.~Hunt, R.~Wilbers, D.~Heyer, C.~de~Kock, H.~Mansvelder, and N.~Goriounova, ``Evolution of cortical neurons supporting human cognition,'' \emph{Trends in Cognitive Sciences}, vol.~26, no.~11, pp. 909--922, 2022.

\bibitem{anCodingCapacityPurkinje2019}
L.~An, Y.~Tang, Q.~Wang, Q.~Pei, R.~Wei, H.~Duan, and J.~K. Liu, ``Coding {{Capacity}} of {{Purkinje Cells With Different Schemes}} of {{Morphological Reduction}},'' \emph{Frontiers in Computational Neuroscience}, vol.~13, 2019.

\bibitem{napper1988number}
R.~Napper and R.~Harvey, ``Number of parallel fiber synapses on an individual purkinje cell in the cerebellum of the rat,'' \emph{Journal of Comparative Neurology}, vol. 274, no.~2, pp. 168--177, 1988.

\bibitem{tangRegulatingSynchronousOscillations2021}
Y.~Tang, L.~An, Q.~Wang, and J.~K. Liu, ``Regulating synchronous oscillations of cerebellar granule cells by different types of inhibition,'' \emph{PLOS Computational Biology}, vol.~17, no.~6, p. e1009163, 2021.

\bibitem{watanabe2011climbing}
M.~Watanabe and M.~Kano, ``Climbing fiber synapse elimination in cerebellar purkinje cells,'' \emph{European Journal of Neuroscience}, vol.~34, no.~10, pp. 1697--1710, 2011.

\bibitem{hodgkinQuantitativeDescriptionMembrane1952}
A.~L. Hodgkin and A.~F. Huxley, ``A quantitative description of membrane current and its application to conduction and excitation in nerve,'' \emph{The Journal of Physiology}, vol. 117, no.~4, pp. 500--544, 1952.

\bibitem{tang2023diverse}
Y.~Tang, X.~Zhang, L.~An, Z.~Yu, and J.~K. Liu, ``Diverse role of {NMDA} receptors for dendritic integration of neural dynamics,'' \emph{PLOS Computational Biology}, vol.~19, no.~4, p. e1011019, 2023.

\bibitem{de1994active}
E.~De~Schutter and J.~M. Bower, ``An active membrane model of the cerebellar purkinje cell. {I}. simulation of current clamps in slice,'' \emph{Journal of neurophysiology}, vol.~71, no.~1, pp. 375--400, 1994.

\bibitem{miyasho2001low}
T.~Miyasho, H.~Takagi, H.~Suzuki, S.~Watanabe, M.~Inoue, Y.~Kudo, and H.~Miyakawa, ``Low-threshold potassium channels and a low-threshold calcium channel regulate {$Ca^{2+}$} spike firing in the dendrites of cerebellar purkinje neurons: a modeling study,'' \emph{Brain research}, vol. 891, no. 1-2, pp. 106--115, 2001.

\bibitem{marasco2013using}
A.~Marasco, A.~Limongiello, and M.~Migliore, ``Using strahler's analysis to reduce up to 200-fold the run time of realistic neuron models,'' \emph{Scientific reports}, vol.~3, no.~1, p. 2934, 2013.

\bibitem{ozawa1998glutamate}
S.~Ozawa, H.~Kamiya, and K.~Tsuzuki, ``Glutamate receptors in the mammalian central nervous system,'' \emph{Progress in neurobiology}, vol.~54, no.~5, pp. 581--618, 1998.

\bibitem{rothman2009synaptic}
J.~S. Rothman, L.~Cathala, V.~Steuber, and R.~A. Silver, ``Synaptic depression enables neuronal gain control,'' \emph{Nature}, vol. 457, no. 7232, pp. 1015--1018, 2009.

\bibitem{zang2021cellular}
Y.~Zang and E.~De~Schutter, ``The cellular electrophysiological properties underlying multiplexed coding in purkinje cells,'' \emph{Journal of Neuroscience}, vol.~41, no.~9, pp. 1850--1863, 2021.

\bibitem{fernandez2024purkinje}
E.~M. Fern{\'a}ndez~Santoro, A.~Karim, P.~Warnaar, C.~I. De~Zeeuw, A.~Badura, and M.~Negrello, ``Purkinje cell models: past, present and future,'' \emph{Frontiers in Computational Neuroscience}, vol.~18, p. 1426653, 2024.

\bibitem{cheron2022electrophysiological}
G.~Cheron, D.~Ristori, J.~Marquez-Ruiz, A.-M. Cebolla, and L.~Ris, ``Electrophysiological alterations of the purkinje cells and deep cerebellar neurons in a mouse model of alzheimer disease (electrophysiology on cerebellum of ad mice),'' \emph{European Journal of Neuroscience}, vol.~56, no.~9, pp. 5547--5563, 2022.

\bibitem{tran2015contribution}
A.~Tran-Van-Minh, R.~D. Caz{\'e}, T.~Abrahamsson, L.~Cathala, B.~S. Gutkin, and D.~A. DiGregorio, ``Contribution of sublinear and supralinear dendritic integration to neuronal computations,'' \emph{Frontiers in cellular neuroscience}, vol.~9, p.~67, 2015.

\bibitem{d2018physiology}
E.~D'Angelo, ``Physiology of the cerebellum,'' \emph{Handbook of clinical neurology}, vol. 154, pp. 85--108, 2018.

\bibitem{magee2000dendritic}
J.~C. Magee, ``Dendritic integration of excitatory synaptic input,'' \emph{Nature Reviews Neuroscience}, vol.~1, no.~3, pp. 181--190, 2000.

\bibitem{schiller2000nmda}
J.~Schiller, G.~Major, H.~J. Koester, and Y.~Schiller, ``{NMDA} spikes in basal dendrites of cortical pyramidal neurons,'' \emph{Nature}, vol. 404, no. 6775, pp. 285--289, 2000.

\bibitem{urban1998active}
N.~N. Urban and G.~Barrionuevo, ``Active summation of excitatory postsynaptic potentials in hippocampal {CA3} pyramidal neurons,'' \emph{Proceedings of the National Academy of Sciences}, vol.~95, no.~19, pp. 11\,450--11\,455, 1998.

\bibitem{magee1999dendritic}
J.~C. Magee, ``Dendritic ih normalizes temporal summation in hippocampal {CA1} neurons,'' \emph{Nature neuroscience}, vol.~2, no.~6, pp. 508--514, 1999.

\bibitem{margulis1998temporal}
M.~Margulis and C.-M. Tang, ``Temporal integration can readily switch between sublinear and supralinear summation,'' \emph{Journal of neurophysiology}, vol.~79, no.~5, pp. 2809--2813, 1998.

\bibitem{de2008high}
C.~de~Solages, G.~Szapiro, N.~Brunel, V.~Hakim, P.~Isope, P.~Buisseret, C.~Rousseau, B.~Barbour, and C.~Lena, ``High-frequency organization and synchrony of activity in the purkinje cell layer of the cerebellum,'' \emph{Neuron}, vol.~58, no.~5, pp. 775--788, 2008.

\bibitem{heck2007beam}
D.~Heck, W.~Thach, and J.~Keating, ``On-beam synchrony in the cerebellum as the mechanism for the timing and coordination of movement,'' \emph{Proceedings of the National Academy of Sciences}, vol. 104, no.~18, pp. 7658--7663, 2007.

\bibitem{person2012purkinje}
A.~L. Person and I.~M. Raman, ``Purkinje neuron synchrony elicits time-locked spiking in the cerebellar nuclei,'' \emph{Nature}, vol. 481, no. 7382, pp. 502--505, 2012.

\bibitem{knogler2019motor}
L.~D. Knogler, A.~M. Kist, and R.~Portugues, ``Motor context dominates output from purkinje cell functional regions during reflexive visuomotor behaviours,'' \emph{Elife}, vol.~8, p. e42138, 2019.

\bibitem{heiney2014precise}
S.~A. Heiney, J.~Kim, G.~J. Augustine, and J.~F. Medina, ``Precise control of movement kinematics by optogenetic inhibition of purkinje cell activity,'' \emph{Journal of Neuroscience}, vol.~34, no.~6, pp. 2321--2330, 2014.

\bibitem{larkum2009synaptic}
M.~E. Larkum, T.~Nevian, M.~Sandler, A.~Polsky, and J.~Schiller, ``Synaptic integration in tuft dendrites of layer 5 pyramidal neurons: a new unifying principle,'' \emph{Science}, vol. 325, no. 5941, pp. 756--760, 2009.

\bibitem{grienberger2014nmda}
C.~Grienberger, X.~Chen, and A.~Konnerth, ``{NMDA} receptor-dependent multidendrite {$Ca^{2+}$} spikes required for hippocampal burst firing in vivo,'' \emph{Neuron}, vol.~81, no.~6, pp. 1274--1281, 2014.

\bibitem{smith2013dendritic}
S.~L. Smith, I.~T. Smith, T.~Branco, and M.~H{\"a}usser, ``Dendritic spikes enhance stimulus selectivity in cortical neurons in vivo,'' \emph{Nature}, vol. 503, no. 7474, pp. 115--120, 2013.

\bibitem{palmer2014nmda}
L.~M. Palmer, A.~S. Shai, J.~E. Reeve, H.~L. Anderson, O.~Paulsen, and M.~E. Larkum, ``{NMDA} spikes enhance action potential generation during sensory input,'' \emph{Nature neuroscience}, vol.~17, no.~3, pp. 383--390, 2014.

\bibitem{takahashi2012locally}
N.~Takahashi, K.~Kitamura, N.~Matsuo, M.~Mayford, M.~Kano, N.~Matsuki, and Y.~Ikegaya, ``Locally synchronized synaptic inputs,'' \emph{Science}, vol. 335, no. 6066, pp. 353--356, 2012.

\bibitem{jia2010dendritic}
H.~Jia, N.~L. Rochefort, X.~Chen, and A.~Konnerth, ``Dendritic organization of sensory input to cortical neurons in vivo,'' \emph{Nature}, vol. 464, no. 7293, pp. 1307--1312, 2010.

\bibitem{chen2011functional}
X.~Chen, U.~Leischner, N.~L. Rochefort, I.~Nelken, and A.~Konnerth, ``Functional mapping of single spines in cortical neurons in vivo,'' \emph{Nature}, vol. 475, no. 7357, pp. 501--505, 2011.

\bibitem{zhou2014cerebellar}
H.~Zhou, Z.~Lin, K.~Voges, C.~Ju, Z.~Gao, L.~W. Bosman, T.~J. Ruigrok, F.~E. Hoebeek, C.~I. De~Zeeuw, and M.~Schonewille, ``Cerebellar modules operate at different frequencies,'' \emph{elife}, vol.~3, p. e02536, 2014.

\bibitem{tang2017heterogeneity}
T.~Tang, J.~Xiao, C.~Y. Suh, A.~Burroughs, N.~L. Cerminara, L.~Jia, S.~P. Marshall, A.~K. Wise, R.~Apps, I.~Sugihara \emph{et~al.}, ``Heterogeneity of purkinje cell simple spike--complex spike interactions: zebrin-and non-zebrin-related variations,'' \emph{The Journal of physiology}, vol. 595, no.~15, pp. 5341--5357, 2017.

\bibitem{caze2012spiking}
R.~Caz{\'e}, M.~Humphries, and B.~Gutkin, ``Spiking and saturating dendrites differentially expand single neuron computation capacity,'' \emph{Advances in neural information processing systems}, vol.~25, 2012.

\bibitem{caze2021all}
R.~D. Caz{\'e}, ``All neurons can perform linearly non-separable computations,'' \emph{F1000Research}, vol.~10, 2021.

\bibitem{gidon2020dendritic}
A.~Gidon, T.~A. Zolnik, P.~Fidzinski, F.~Bolduan, A.~Papoutsi, P.~Poirazi, M.~Holtkamp, I.~Vida, and M.~E. Larkum, ``Dendritic action potentials and computation in human layer 2/3 cortical neurons,'' \emph{Science}, vol. 367, no. 6473, pp. 83--87, 2020.

\bibitem{masoli2023human}
S.~Masoli, D.~Sanchez-Ponce, N.~Vrieler, K.~Abu-Haya, V.~Lerner, T.~Shahar, H.~Nedelescu, M.~F. Rizza, R.~Benavides-Piccione, J.~DeFelipe \emph{et~al.}, ``Human outperform mouse purkinje cells in dendritic complexity and computational capacity,'' \emph{bioRxiv}, pp. 2023--03, 2023.

\bibitem{busch2023climbing}
S.~E. Busch and C.~Hansel, ``Climbing fiber multi-innervation of mouse purkinje dendrites with arborization common to human,'' \emph{Science}, vol. 381, no. 6656, pp. 420--427, 2023.

\bibitem{najafi2013beyond}
F.~Najafi and J.~F. Medina, ``Beyond “all-or-nothing” climbing fibers: graded representation of teaching signals in purkinje cells,'' \emph{Frontiers in neural circuits}, vol.~7, p. 115, 2013.

\bibitem{steuber2007cerebellar}
V.~Steuber, W.~Mittmann, F.~E. Hoebeek, R.~A. Silver, C.~I. De~Zeeuw, M.~H{\"a}usser, and E.~De~Schutter, ``Cerebellar {LTD} and pattern recognition by purkinje cells,'' \emph{Neuron}, vol.~54, no.~1, pp. 121--136, 2007.

\bibitem{majoral2020model}
D.~Majoral, A.~Zemmar, and R.~Vicente, ``A model for time interval learning in the purkinje cell,'' \emph{PLoS computational biology}, vol.~16, no.~2, p. e1007601, 2020.

\bibitem{zang2018voltage}
Y.~Zang, S.~Dieudonn{\'e}, and E.~De~Schutter, ``Voltage-and branch-specific climbing fiber responses in purkinje cells,'' \emph{Cell reports}, vol.~24, no.~6, pp. 1536--1549, 2018.

\bibitem{guan2020minimal}
B.~Guan, J.~Zhao, Z.~Li, F.~Sun, and F.~Fraundorfer, ``Minimal solutions for relative pose with a single affine correspondence,'' in \emph{Proceedings of the IEEE/CVF Conference on Computer Vision and Pattern Recognition}, 2020, pp. 1929--1938.

\bibitem{guan2021relative}
------, ``Relative pose estimation with a single affine correspondence,'' \emph{IEEE Transactions on Cybernetics}, vol.~52, no.~10, pp. 10\,111--10\,122, 2021.

\bibitem{tamura2023discrimination}
K.~Tamura, Y.~Yamamoto, T.~Kobayashi, R.~Kuriyama, and T.~Yamazaki, ``Discrimination and learning of temporal input sequences in a cerebellar purkinje cell model,'' \emph{Frontiers in Cellular Neuroscience}, vol.~17, p. 1075005, 2023.

\bibitem{masoli2024human}
S.~Masoli, D.~Sanchez-Ponce, N.~Vrieler, K.~Abu-Haya, V.~Lerner, T.~Shahar, H.~Nedelescu, M.~F. Rizza, R.~Benavides-Piccione, J.~DeFelipe \emph{et~al.}, ``Human purkinje cells outperform mouse purkinje cells in dendritic complexity and computational capacity,'' \emph{Communications Biology}, vol.~7, no.~1, p.~5, 2024.

\bibitem{caze2024demonstration}
R.~D. Caz{\'e}, A.~Tran-Van-Minh, B.~S. Gutkin, and D.~A. DiGregorio, ``Demonstration that sublinear dendrites enable linearly non-separable computations,'' \emph{Scientific Reports}, vol.~14, no.~1, p. 18226, 2024.

\bibitem{guan2022affine}
B.~Guan and J.~Zhao, ``Affine correspondences between multi-camera systems for 6dof relative pose estimation,'' in \emph{European Conference on Computer Vision}.\hskip 1em plus 0.5em minus 0.4em\relax Springer, 2022, pp. 634--650.

\bibitem{yu2023binding}
X.~Yu and E.~Lau, ``The binding problem 2.0: beyond perceptual features,'' \emph{Cognitive Science}, vol.~47, no.~2, p. e13244, 2023.

\bibitem{von1999and}
C.~Von~der Malsburg, ``The what and why of binding: the modeler’s perspective,'' \emph{Neuron}, vol.~24, no.~1, pp. 95--104, 1999.

\bibitem{sun2025learning}
P.~Sun, B.~Guan, Z.~Yu, Y.~Shang, Q.~Yu, and D.~Barath, ``Learning affine correspondences by integrating geometric constraints,'' \emph{arXiv preprint arXiv:2504.04834}, 2025.

\bibitem{ito2008control}
M.~Ito, ``Control of mental activities by internal models in the cerebellum,'' \emph{Nature Reviews Neuroscience}, vol.~9, no.~4, pp. 304--313, 2008.

\bibitem{ujfalussy2018global}
B.~B. Ujfalussy, J.~K. Makara, M.~Lengyel, and T.~Branco, ``Global and multiplexed dendritic computations under in vivo-like conditions,'' \emph{Neuron}, vol. 100, no.~3, pp. 579--592, 2018.

\bibitem{tang2021modulation}
Y.~Tang, L.~An, Y.~Yuan, Q.~Pei, Q.~Wang, and J.~K. Liu, ``Modulation of the dynamics of cerebellar purkinje cells through the interaction of excitatory and inhibitory feedforward pathways,'' \emph{PLoS computational biology}, vol.~17, no.~2, p. e1008670, 2021.

\bibitem{pouzat1997developmental}
C.~Pouzat and S.~Hestrin, ``Developmental regulation of basket/stellate cell→ purkinje cell synapses in the cerebellum,'' \emph{Journal of Neuroscience}, vol.~17, no.~23, pp. 9104--9112, 1997.

\bibitem{dizon2011role}
M.~J. Dizon and K.~Khodakhah, ``The role of interneurons in shaping purkinje cell responses in the cerebellar cortex,'' \emph{Journal of Neuroscience}, vol.~31, no.~29, pp. 10\,463--10\,473, 2011.

\bibitem{dezeeuwDiversityDynamismCerebellum2021}
C.~I. De~Zeeuw, S.~G. Lisberger, and J.~L. Raymond, ``Diversity and dynamism in the cerebellum,'' \emph{Nature Neuroscience}, vol.~24, no.~2, pp. 160--167, 2021.

\bibitem{zhuActivityMapCorticocerebellar2023}
J.~Zhu, H.~Hasanbegović, L.~D. Liu, Z.~Gao, and N.~Li, ``Activity map of a cortico-cerebellar loop underlying motor planning,'' \emph{Nature Neuroscience}, vol.~26, no.~11, pp. 1916--1928, 2023.

\end{thebibliography}

\newpage
\onecolumn
\begin{center}
\Large{Supplemental Materials:} 
\hspace{1cm}
\hspace{2cm}
\end{center}
\renewcommand{\thefigure}{S\arabic{figure}}
\setcounter{figure}{0}
\renewcommand{\thetable}{S\arabic{table}}
\setcounter{table}{0}
\hspace{2cm}

\begin{figure*}[hbp]
	\centering	
    \includegraphics[width=0.9\textwidth]{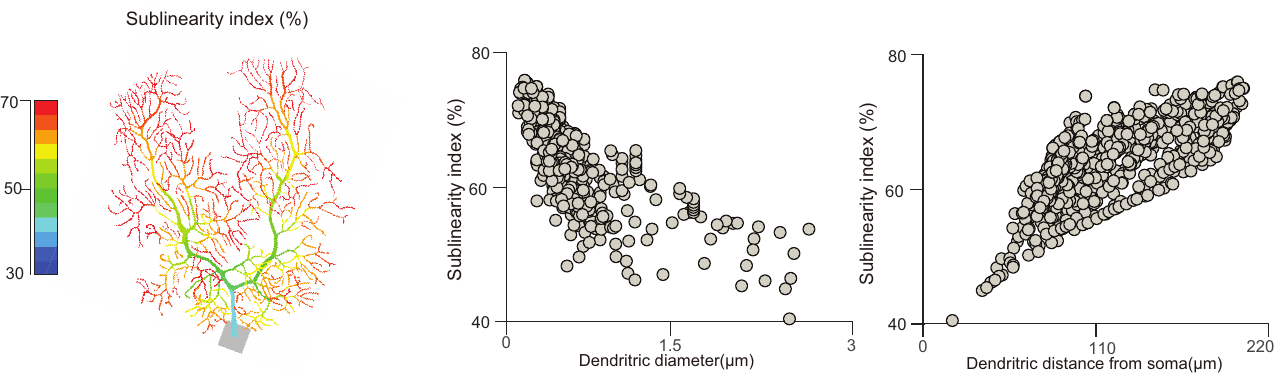}
	\caption{  Related to Figure~\ref{fig1}.
    The sublinearity index (\%) across the dendritic tree was analyzed in a PC model that incorporates active ion channels. The color-coded morphology (left) illustrates the variation in sublinearity across the dendritic tree. The scatter plots depict the sublinearity index as a function of dendritic diameter (middle) and dendritic distance from the soma (right).} 
    
	\centering
	\label{suppfig1}
\end{figure*}

\begin{figure*}[hbp]
	\centering	
    \includegraphics[width=0.9\textwidth]{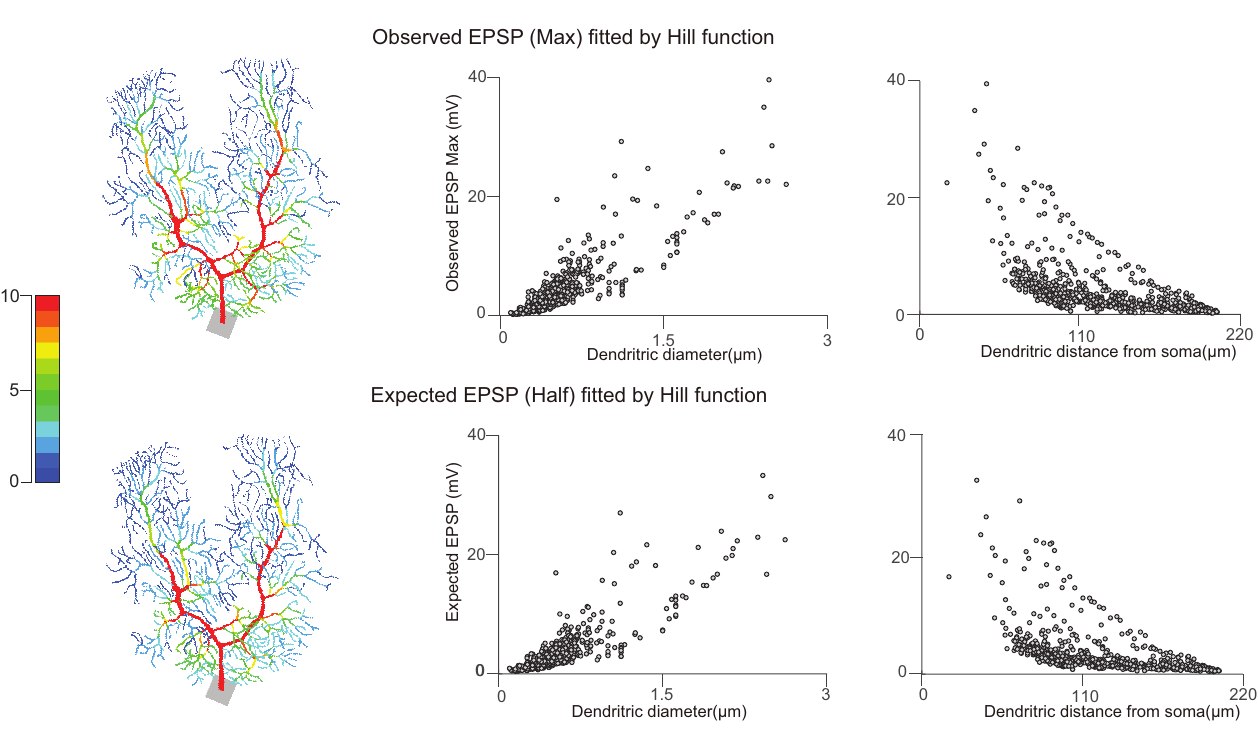}
	\caption{ Related to Figure~\ref{fig1}. Color-coded (Top) the maximum observed EPSP (Max) and (Bottom) the value of expected EPSP at which observed EPSP reaches half maximum (Half) obtained by Hill function fits. Max and Half EPSP (Observed: Top, Expected: Bottom) as a function of dendritic diameter and distance from soma.}
	\centering
	\label{suppfig2}
\end{figure*}

\begin{figure*}[hbp]
	\centering	
    \includegraphics[width=0.9\textwidth]{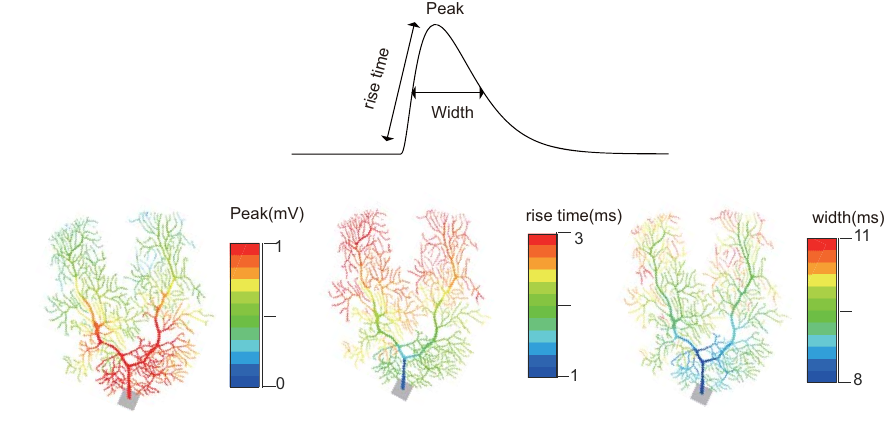}
	\caption{ Related to Figure~\ref{fig1}. The top panel illustrates the key parameters of an electrical signal: peak amplitude, rise time, and width of EPSP when a single synapse is distributed on different dendrites. The bottom panels display heatmaps of dendritic branches, visualizing the distribution of these parameters. From left to right, the maps show: Peak amplitude (mV), Rise time (ms) and width (ms). Color scales indicate the value ranges for each parameter, with red representing higher values and blue representing lower values. This visualization highlights the spatial variability of signal properties across the dendritic tree.
 }
	\centering
	\label{suppfig3}
\end{figure*}

\begin{figure*}[bp]
	\centering	
    \includegraphics[width=0.9\textwidth]{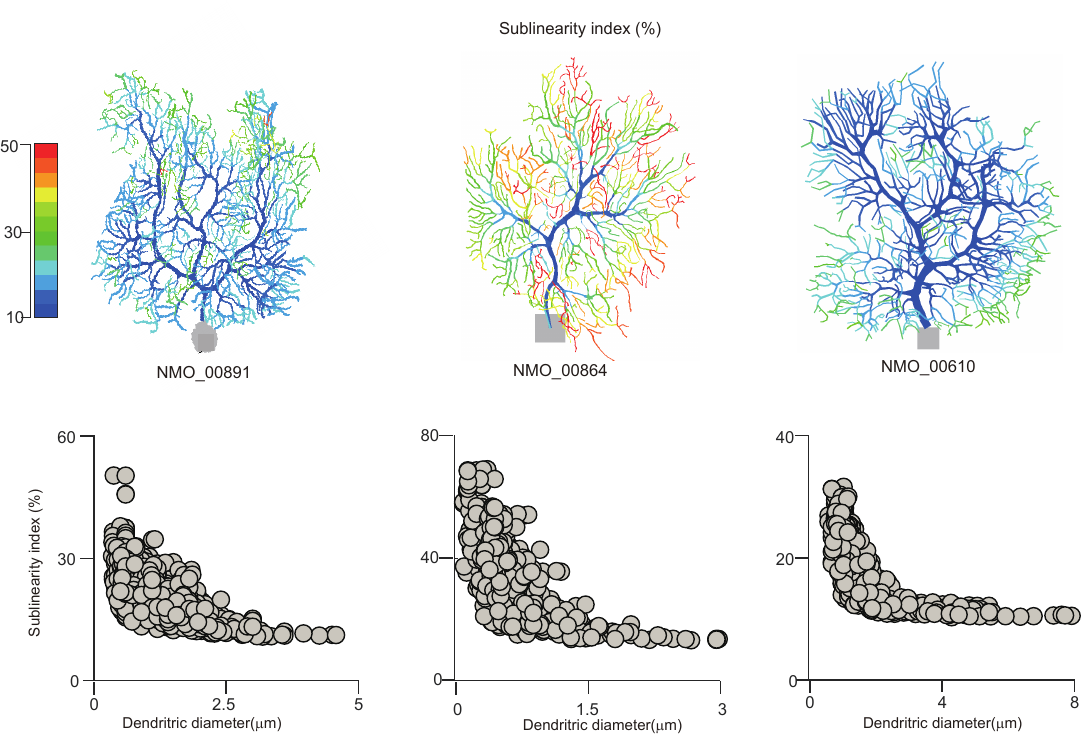}
	\caption{ Related to Figure~\ref{fig1}. The distributed sublinearity index of each dendrite from three additional PCs. (Left) A rat Purkinje cell (NMO-00891). (Middle) A mouse Purkinje cell (NMO-00864). (Right) A guinea pig Purkinje cell (NMO-00610). The morphology structures were taken from the public archive (www.neuromorpho.org).}
	
	\centering
	\label{suppfig4}
\end{figure*}

\begin{figure*}[hbp]
	\centering	
    \includegraphics[width=0.9\textwidth]{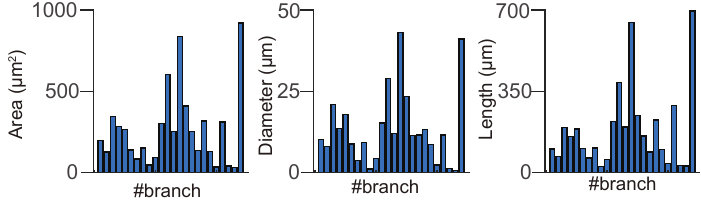}
	\caption{Related to Figure\ref{fig3}. Histograms show the distribution of characteristics, (left) surface area, (middle) diameter, and (right) length, of 24 branches in the neuron model. }
 
	\centering
	\label{suppfig5}
\end{figure*}

% \begin{figure*}[bp]
% 	\centering	
%     \includegraphics[width=0.8\textwidth]{SF3.pdf}
% 	\caption{ Related to Figure\ref{fig4}. Left column: Diagrams of three synaptic input distribution patterns: Cluster Input: Synapses concentrated in a single dendrite. Local Scattered: Synapses distributed across a branch area. Global Scattered: Synapses spread across the entire dendritic tree. Middle column: Output spike frequency (Hz) as a function of input spike frequency (Hz) with different numbers of synapses (50: red, 100: green, 150: blue). Right column: Output spike frequency (Hz) as a function of input spike frequency (Hz) with different input rate (50 Hz: red, 100 Hz: green, 150 Hz: blue). Here is synchronous input.
% 	}
% 	\centering
% 	\label{suppfig3}
% \end{figure*}

\begin{figure*}[bp]
	\centering	
    \includegraphics[width=0.9\textwidth]{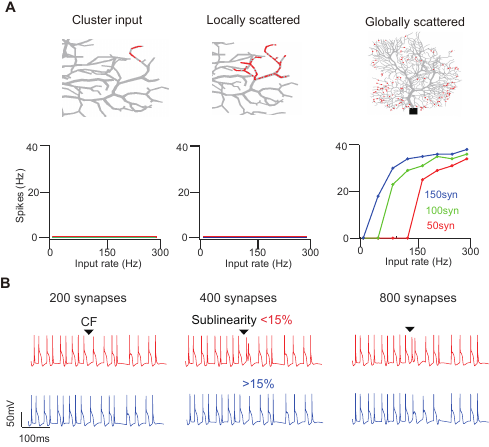}
	\caption{ Related to Figure~\ref{fig4} and Figure~\ref{fig6}. The Purkinje cell was modeled from guinea pig (NMO\_00610). 
    (A). Spiking activity vs. asynchronous input under three spatial stimulation
patterns with 50, 100, and 150 synapses. 
    (B). Strong sublinearity suppresses burst. PC installed with 1000 PF AMPA synapses stimulated
at 30 Hz Poisson spikes and a CF with 200, 400, and 800 AMPA+NMDA synapses distributed
on different sublinear dendrites. }
	
	\centering
	\label{suppfig6}
\end{figure*}

\begin{figure*}[bp]
	\centering	
    \includegraphics[width=0.9\textwidth]{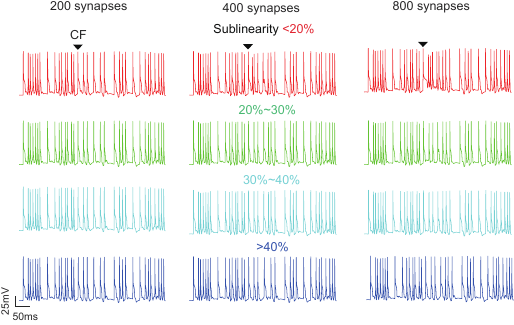}
	\caption{ Related to Figure~\ref{fig6}. Strong sublinearity suppresses burst. PC installed with 1000 PF AMPA synapses stimulated at 100 Hz Poisson spikes and a CF with 200, 400 and 800 AMPA+NMDA synapses distributed on different sublinear dendrites.}
	
	\centering 
	\label{suppfig7}
\end{figure*}

\begin{figure*}[tbp]
	\centering	
    \includegraphics[width=0.9\textwidth]{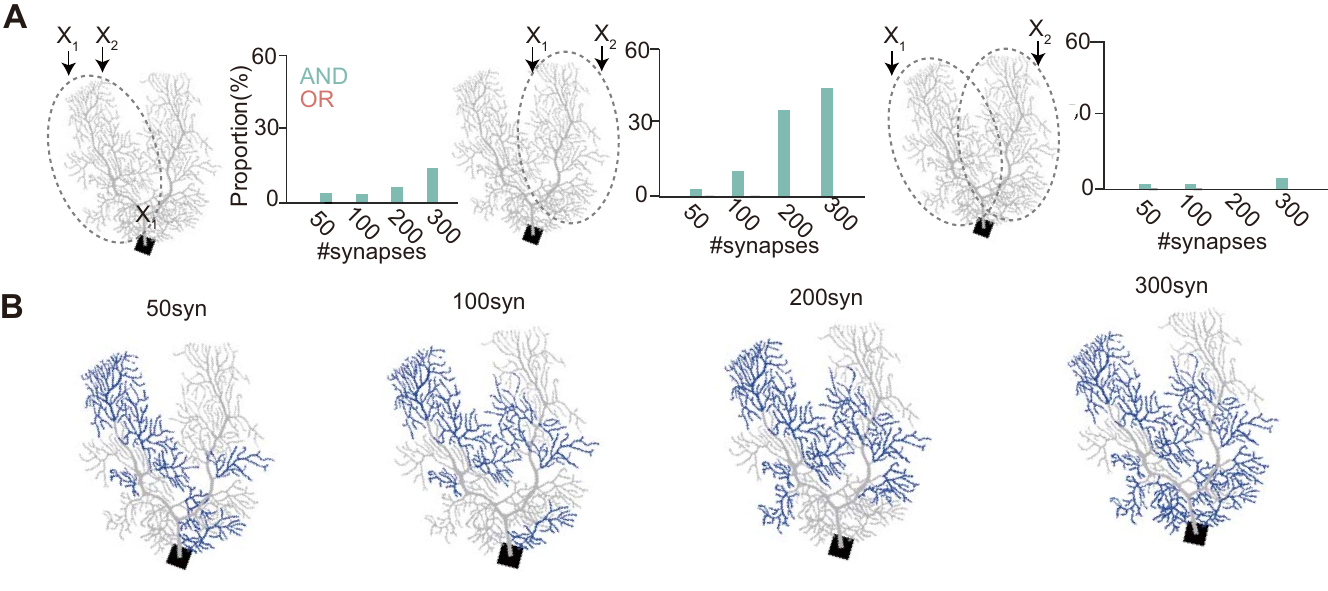}
	\caption{ Related to Figure~\ref{fig7}. (A) The proportion of AND and OR computations with varying numbers of synapses (50, 100, 200, 300) distributed on the left, right, and both the main branches with asynchronous inputs at 50Hz. (B) The distribution of branches participating in AND (blue) and OR (green) with different
synapse inputs (50, 100, 200, 300). Gray areas indicate regions not participating in any Boolean operations. 
	}
	\centering
	\label{suppfig8}
\end{figure*}

\begin{figure*}[tbp]
	\centering	
    \includegraphics[width=0.9\textwidth]{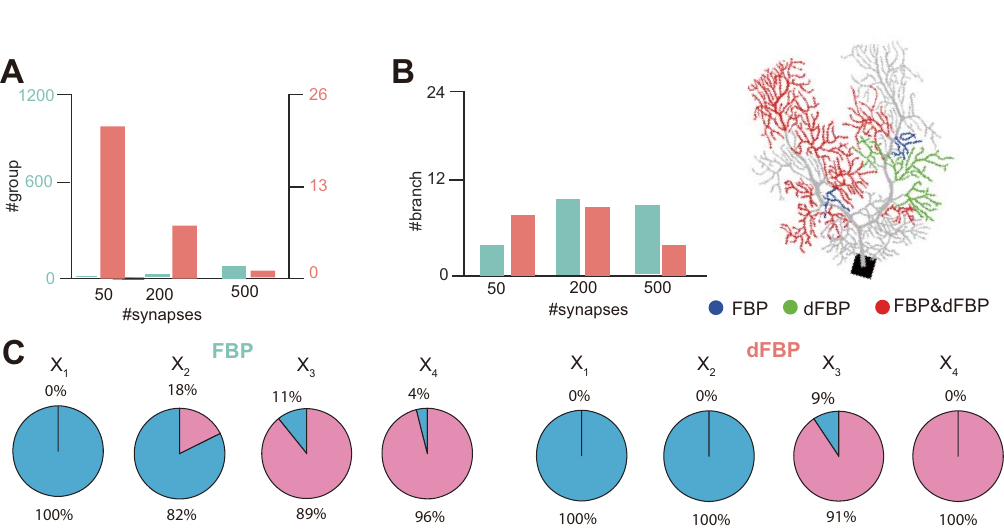}
	\caption{ Related to Figure~\ref{fig8}. (A) The number of branch ensembles (groups) for FBP and dFBP with an input of different numbers of synapses (20, 200, 500) with synchronous input at 50Hz.
   (B) The number of branches for FBP and dFBP and their distribution on the dendritic tree. Blue regions indicate participation only in feature binding problem (FBP), green regions indicate participation only in double feature binding problem (dFBP), and red regions indicate participation in both FBP and dFBP. Gray areas indicate regions not participating in any feature binding operations. 
   (C) The distribution of 4 input vectors on the left and right main branches when implementing FBP and dFBP. 
	}
	\centering
	\label{suppfig9}
\end{figure*}

\begin{figure*}[tbp]
	\centering	
    \includegraphics[width=0.9\textwidth]{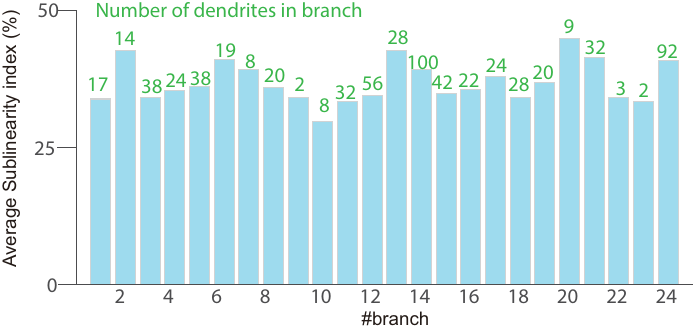}
	\caption{ Related to Figure~\ref{fig8}. The bar chart shows the average sublinearity index (\%) for each dendritic branch. The green numbers above each bar indicate the number of dendrites in each branch. The sublinearity index reflects the degree of sublinear integration occurring in each branch, with higher values indicating greater sublinear integration. This data illustrates the variability in sublinear integration across different dendritic branches.
	}
	\centering
	\label{suppfig10}
\end{figure*}

 \begin{figure*}[tbp]
	\centering	
    \includegraphics[width=0.9\textwidth]{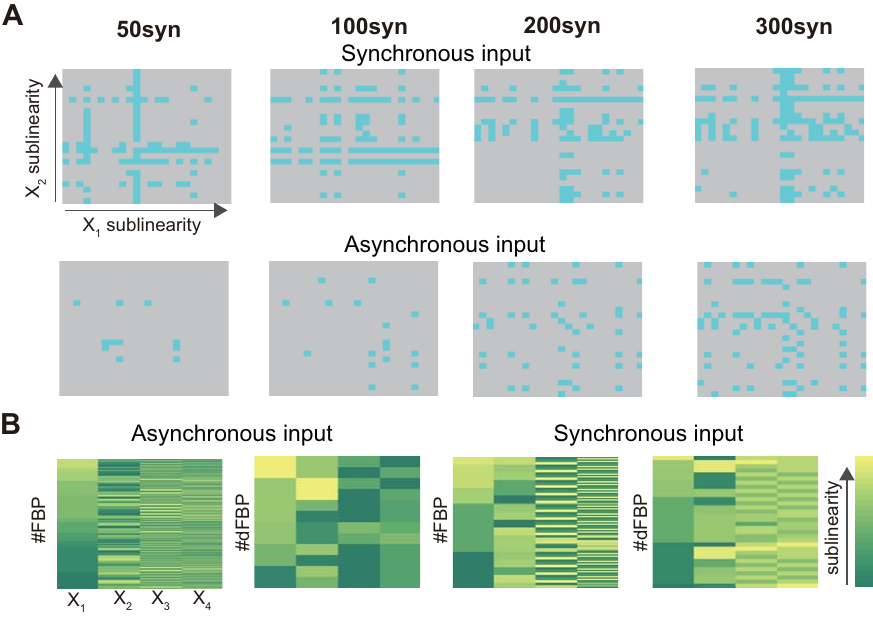}
    \caption{ Related to Figure~\ref{fig8}.
    (A) Schematic diagram illustrating the sorting of dendritic branch sublinearity indices (SI) and the implementation of AND operations (cyan) with different numbers of synapses (50, 100, 200, 300) under synchronous and asynchronous input conditions. In general, it is not easy to produce AND operations when the sublinearities of x1 and x2 are both weak. Higher sublinearity indices facilitate AND operations. (B)  When executing FBP and dFBP, a schematic diagram of the sublinearity of the four branches. } 
	\centering
	\label{suppfig11}
\end{figure*}

\end{document}